Acta Appl.Math.  28, N1,(1992), 1-42.
\def\aaa{A(\alpha ',\alpha )}
\def\adaa{A_{\delta }(\alpha ',\alpha )}
\def\sqr#1#2{{\vcenter{\vbox {\hrule height.#2pt
\hbox{\vrule width.#2pt height#1pt \kern#1pt \vrule width.#2pt}
\hrule height.#2pt}}}}
\def\square{\mathchoice \sqr34 \sqr34 \sqr{2.1}3\sqr{1.5}3}
\par \centerline {\bf STABILITY ESTIMATES IN INVERSE SCATTERING}
\par \centerline {A. G. Ramm, Mathematics Department,}
\par \centerline {Kansas State University, Manhattan, KS 66506-2602,
USA}
\footnote{}{Key words and phrases: Inverse scattering,
stability, noisy data}
\footnote{}{AMS Mathematics subject classification 35R30 (1991 Revision)}
\par \noindent {{\bf Abstract}: An algorithm is given for calculating
the solution to the 3D inverse scattering problem with noisy discrete
fixed energy data. The error estimates for the calculated solution are
derived. The methods developed are of general nature and can be used in
many applications: in nondestructive evaluation and remote sensing, in
geophysical exploration, medical diagnostics and technology.}  \vskip 1pc
\par \centerline {\bf I. Introduction}
\par {Let $q\in Q:=\{q\colon q(x)=0$ for $|x|\geq a, \quad x\in R^3,
\quad q(x)={\bar q}(x), \quad q\in L^{\infty }\},\quad a>0$ is
a constant and the bar stands for complex conjugate. Consider the
equation
  $$ l_qu:=[\nabla ^2+1-q(x)]u(x,\alpha )=0 \hbox { in } R^3,\quad
     \alpha \in S^2  \eqno (1.1) $$
where $S^2$ is the unit sphere in $R^3$ and
  $$ u(x,\alpha )=\exp (i\alpha \cdot x)+r^{-1}\exp (ir)\aaa +o(r^{-1}),
  \qquad r=|x|\to \infty ,\quad \alpha '=x/r \eqno (1.2) $$
 The existence and uniqueness of the solution to (1), (2) are well known
and are easy to prove (see e.g. [1,2]). The coefficient
$A(\alpha ',\alpha )$ is called the scattering amplitude at the fixed
energy ($k^2=1$). Its properties are discussed e.g. in [2], pp.
219--246. The {\it inverse scattering problem} (ISP) consists of
finding $q\in Q$ from the knowledge of $\aaa $ for all
$\alpha ',\alpha \in S^2$. Uniqueness of the solution to ISP is proved
in [3,4], (see also [2]) even in the case when $\aaa $ is known
for $\alpha \in \tilde {S_1^2}$, $\alpha '\in \tilde {S_2^2}$, where
$\tilde {S_j^2}$, $j=1,2,$ are arbitrary small open sets in $S^2$.}
\par {The uniqueness result holds with the same proof as in [4] for
the problem
  $$ [\nabla ^2+1-q_0(x)-q(x)]u=0 \hbox { in } R^3
  \hbox { and (1.2) holds } \eqno (1.1') $$
where $q_0(x)$ is a {\it known} short-range potential, e.g.
$|q_0(x)|\leq c(1+|x|)^{-2-\varepsilon }$,
$\varepsilon >0$ and $q(x)\in Q$
(this was noticed by R. Weder, in a preprint, 1991). The aim
of this paper is to complete the study of the stability of the solution
to ISP with noisy data which the author carried over in the papers
[4--18] and in the monograph [1]. The statement of this problem
(${\rm ISP}_\delta $) is as follows.}
\par {Let a function $\adaa $ be given which is not assumed to be a
scattering amplitude corresponding to a potential and let
  $$ \mathop {\sup }_{\alpha ',\alpha \in S^2}|\adaa -\aaa |<\delta
     \eqno (1.3) $$
The inverse scattering problem with noisy data is:}
\par \noindent {(${\rm ISP}_\delta $): {\it Given $\adaa $ for all
$\alpha ', \alpha \in S^2$ find ${\hat q}_\delta (x)$ such that}
  $$ \sup_{\lambda \in R^3}|\hat q_\delta -\tilde q(\lambda )|
  <\eta (\delta )\to 0 \hbox { as } \delta \to 0 \eqno (1.4) $$
{\it where}
$$ \tilde q(\lambda ):=\int _{R^3}\exp (-i\lambda \cdot x)q(x)\,dx. $$
In other words, we want to calculate a stable approximation to
$\tilde q(\lambda )$ and estimate the rate of decay of $\eta (\delta )$
as $\delta \to 0$. To get such an estimate uniform in $q$ belonging to
a compact subset of $Q$ we assume that
  $$ q\in {\cal B}_C:=\{q\colon \Vert q \Vert _{L^{\infty }} +
      \Vert \nabla q\Vert _{L^{\infty }} \leq C\} \eqno (1.5) $$
The ${\rm ISP}_\delta  $ is of considerable practical interest: it is a
theoretical basis for many problems in nondestructive evaluation and
remote sensing, including, to mention a few, problems of geophysical
exploration, medical diagnostics, technology, etc. To the author's
knowledge, there were no results concerning three-dimensional
${\rm ISP}_\delta $ except those given in the works [1,4--18]}
\par {The basic results of this paper are new. They are formulated as
Theorems A, B and C in Section II. Theorem A gives a rigorous inversion
procedure for ISP with exact data and an error estimate for this
procedure. Theorem B gives a rigorous contruction of the solution to
${\rm ISP}_\delta $, and an estimate for the difference between this
solution and the unknown potential. Theorem B is of practical interest
and can be used in designing a numerical code for
solving ${\rm ISP}_\delta $. An alternative method is given in Section
VII.3.   Theorem C gives an estimate for the difference of the Fourier
transforms of two potentials if an estimate for the difference of the
corresponding scattering amplitudes is given at a fixed energy.}
\par {There are other new results in this paper. They are formulated
as lemmas used in the proof of the basic results. Although the ideas
and the techniques from works [4--18] are used in this paper there are
many improvements, simplifications in the arguments and the paper is
essentially self-contained: no prior knowledge of the author's work is
assumed. The results may be of interest to mathematicians, engineers,
physicists and numerical analysts interested in ${\rm ISP}_\delta $.}
\par {The paper is organized as follows: Section II contains the
statement of the basic results and a number lemmas. Section III
contains auxiliary results and proofs of Theorem A and lemmas 10
through 17. Section IV contains proofs of lemmas 1 through 9 except
lemma 6, which is proved in Section V. In Section VI proofs of Theorems
B and C are given. In Section VII some numerical aspects of the
${\rm ISP}_\delta $ are discussed and a summary of the numerical
approach is given. In addition, an alternative numerical method for
solving ${\rm ISP}_\delta $ is discussed and an error estimate for
this method is proved.}
\par \centerline {\bf II Formulation of the results}
\par {1. First, assume that the exact data are given:
  $$ \aaa =\sum _{\ell =0}^\infty A_{\ell }(\alpha )Y_\ell
     (\alpha '), \qquad  A_\ell (\alpha):=\int \limits_{S^2}
     \aaa {\bar Y_\ell (\alpha ')}\,d\alpha '\eqno (2.1) $$
Here and below the summation in $\ell $ denotes
  $ \sum_{\ell =0}^\infty \sum_{m=-\ell }^\ell ,\quad
  A_\ell (\alpha )=A_{\ell m}(\alpha ),\quad
  Y_\ell (\alpha )=Y_{\ell m}(\alpha )$,\break $-\ell \leq m\leq \ell $,
  $$ Y_{\ell m}(\alpha )={1\over \sqrt{ 4\pi} }\left [
  {(2\ell +1)(\ell -|m|)! \over (\ell +|m|)!}\right ]^{1/2}
  P_{\ell ,|m|}(\cos \vartheta ) \exp (im\phi ),  \eqno (2.2) $$
where $(\vartheta ,\phi )$ are the angles in the spherical coordinates
determining the unit vector $\alpha \in S^2$:
  $$ \alpha _1=\sin \vartheta \cos \phi , \qquad \alpha _2=\sin
  \vartheta \sin \phi , \qquad \alpha _3=\cos \vartheta \eqno (2.3) $$
Define
  $$ M:=\{\theta \colon \theta \in C^3, \quad
     \theta \cdot \theta =1\}, \eqno (2.4) $$
 
  $$ \theta \cdot \omega =\theta_1\cdot \omega_1+\theta_2\cdot \omega_2
     +\theta_3\cdot \omega_3, \qquad |\theta |=(\theta \cdot \bar \theta)
     ^{1/2} \eqno (2.5) $$
If $0\leq \vartheta \leq \pi,\quad 0\leq \phi <2\pi $, then the vector
$\alpha =(\alpha _1,\alpha _2,\alpha _3)$ runs through $S^2$. If
$\vartheta ,\phi $ run through the complex plane $C$ then the
corresponding $\alpha $ runs through $M'\subset M$. 
The subset $M'$ contains all the vectors of $M$ except
the ones of the form
 $ (v_1,v_2,1)$ and $(v_1,v_2,-1)$, where $v:=(v_1, v_2)\in C^2$ are
vectors with the property $v\cdot v=0$, and $v\neq 0$.
From (2.2) it follows that
$Y_{\ell m}(\alpha )$ is defined for all $\alpha \in M'$: it is
sufficient to find for $\alpha \in M'$ the corresponding complex numbers
$\vartheta $ and $\phi $ such that formulas (2.3) hold and calculate the
right-hand side of (2.2) for these $\vartheta $ and $\phi $. Obviously
$\exp (im\phi )$ is defined for complex $\phi $, and the function
  $$ P_{\ell ,|m|}(\cos \vartheta )=(\sin \vartheta )^{|m|}{d^{|m|}
P_{\ell } (\cos \vartheta )\over (d\,\cos \vartheta )^m} \eqno (2.6) $$
is defined for complex $\vartheta $. Let
  $$ B_a:=\{x\colon x\in R^3,\quad |x|\leq a\}, \qquad
     Q_a:=\{q\colon q=\bar q,\quad q=0
     \hbox{ for } |x|>a,\quad q\in L^2(B_a)\}.$$
{\bf Lemma 1}: {\sl If $q\in Q_a$ then}
  $$ \mathop {\sup_{\alpha \in S^2 }}_{-\ell \leq m\leq \ell }
  |A_{\ell m}(\alpha )|\leq ca\left ({ae\over 2\ell +1} \right )
  ^{2\ell +1\over 2}{1\over 2\ell +1}  \eqno (2.7) $$
{\bf Lemma 2}: {\sl If $\theta \in M'$ then}
  $$ \vert Y_{\ell }(\theta )\vert \leq {1\over \sqrt {4\pi }}
     {\exp (\kappa r)\over |j_{\ell }(r) |},\qquad r>0, \qquad
     \kappa :=|\hbox{Im }\theta | \eqno (2.8) $$
{\sl where $r>0$ is an arbitrary fixed number and $j_{\ell }(r)$ is the
spherical Bessel function,}}
\par \noindent {$ j_\ell (r):=[\pi /(2r)]^{1/2}
  J_{\ell +{1\over 2}}(r), \quad J_{\ell +{1\over 2}}(r) \ $
  {\sl is the usual Bessel function.}
\par {It is well known that
  $$ j_\ell (r)={1\over \sqrt {2r}}\left ({er\over 2\ell +1}\right )
  ^{2\ell +1\over 2}{1\over \sqrt {2\ell +1}}[1+o(1)] \hbox{ as }
   l\to \infty \eqno (2.9) $$
uniformly in $r\in [0,a]$ for any fixed $a>0$. From (2.7)--(2.9) it
follows that the series
  $$ A(\theta ',\alpha )=\sum _{\ell =0}^\infty A_\ell (\alpha )
  Y_\ell (\theta '), \qquad \theta '\in M, \quad \alpha \in S^2
  \eqno (2.10) $$
converges absolutely and uniformly on $\tilde M\times S^2$, where
$\tilde M\subset M'$ is an arbitrary compact subset in $M'$. Indeed,
choose $r>a$ in (2.8). Then the series (2.10) is majorized by the
convergent series
  $$ c\,\exp (\kappa r)\sum_{\ell =0}^\infty \sqrt {2\ell +1}
     \left ({a\over r}\right )^{2\ell +1\over 2}  \eqno (2.11) $$
Note that given the data $\aaa $ $\forall \alpha ',\alpha \in S^2$
one finds the Fourier coefficients $A_\ell (\alpha )$ by formula
(2.1) and defines $A(\theta ',\alpha )$
$\forall \theta '\in M, \, \alpha \in S^2$ by formula (2.10).}
\par {2. We now pass to the description of the inversion
formula for exact data. Define
  $$ \rho (\nu ):=\exp (-i\theta \cdot x)\int \limits_{S^2}u(x,\alpha )
  \nu (\alpha )\,d\alpha -1, \qquad \theta \in M',\quad
  \nu \in L^2(S^2) \eqno (2.12) $$
Consider the variational problem
  $$ \Vert \rho (\nu ) \Vert :=\inf :=d(\theta ) \eqno (2.13) $$
where $\rho (\nu )$ is a function of $x\in R^3 $ and $\theta\in M'$,
  $$ \Vert \rho \Vert :=
  \Vert \rho \Vert _{L^2(B_b\setminus B_{a_1})}, \qquad
     a<a_1<b \eqno (2.14) $$
and $a_1,b$ are arbitrary numbers subject to the restriction (2.14).
Note that in the annulus $a_1<|x|<b$ the scattering solution
$u(x,\alpha )$ is defined explicitly through the data. Namely the
following simple lemma holds:}
\par \noindent {{\bf Lemma 3}: {\sl One has}
  $$ u(x,\alpha )=\exp (i\alpha \cdot x)+\sum_{\ell =0}^\infty
  A_\ell (\alpha )Y_\ell (\alpha ')h_\ell (r),\quad r>a,\quad
  r=|x|,\quad \alpha '=x/r \eqno (2.15) $$
{\sl where $h_\ell (r)$ are the spherical Hankel functions normalized
so that  $h_\ell (r)\sim r^{-1}\exp (ir)$ as $r\to \infty $.}}
\par \noindent {{\bf Lemma 4}: {\sl If $q\in Q_a$ then}
  $$ d(\theta )\leq c|\theta |^{-1}, \qquad |\theta | \gg 1,
     \quad \theta \in M'   \eqno (2.16) $$}
\par {Here and below $c$ stand for various positive constants which do
not depend on $\theta $ and $\delta $. These constants depend on the
norm $\Vert q\Vert _a:=\Vert q\Vert _{L^2(B_a)}$ for $q\in Q$, and for
$q\in {\cal B}_C$ they depend on $C$ where $C$ is defined in (1.5). The
notation $|\theta |\gg 1$ means that $\theta $ is sufficiently large.}
\par {Denote by $\nu (\alpha ,\theta )$ an arbitrary function in
$L^2(S^2)$ which satisfies the inequality
  $$ \Vert \rho \bigl (\nu (\alpha ,\theta )\bigr )\Vert \leq
  (c+1)|\theta |^{-1},  \qquad |\theta |\to \infty \eqno (2.17) $$
It follows from Lemma 4 that such a function can be calculated by solving
problem (2.13). Define for this $\nu (\alpha ,\theta )$ the quantity
  $$ \hat q:=-4\pi \int \limits_{S^2}A(\theta ',\alpha )\nu (\alpha ,
  \theta )\, d\alpha ,\qquad \theta ',\theta \in M' \eqno (2.18)  $$
Fix an arbitrary large number $\lambda _0>0$, take any $\lambda\in R^3$,
$|\lambda |\leq \lambda _0$, and pick any $\theta ',\theta $ such that
  $$ \theta '-\theta =\lambda ,\quad \theta',\theta \in M, \quad
     |\theta |\to \infty \eqno (2.19) $$
{\bf Lemma 5}: {\sl For any $\lambda \in R^n$, $n\geq 3$ there exist
$\theta ,\theta '\in M$ satisfying (2.19).}}
\par \noindent {{\bf Remark:} For $n=2$ this is not true as one can
easily check. If $\lambda_3\neq \pm 1$ then one can find
$\theta ,\theta '\in M'$ satisfying (2.19). In what follows we may
assume without loss of generality that $\lambda_3\neq \pm 1$
and use $M'$ everywhere. Indeed, we can find the Fourier
transform of the potential, $\tilde q(\lambda)$ for any
$\lambda$ with $\lambda_3 \neq \pm 1$ and since
$\tilde q$ is continuous, in fact analytic, with respect to
$\lambda$, the value of $\tilde q$ at $\lambda=(0,0, \pm 1)$
is determined uniquely by the continuity of $\tilde q$.}
\par {Let
  $$ \tilde q(\lambda ):=\int \limits_{B_a}q(x)
     \exp (-i\lambda \cdot x)\, dx \eqno (2.20) $$ }
\par {3. We are ready to formulate the first theorem.}
\par \noindent {{\bf Theorem A}: {\sl If $q\in Q$, $\hat q$ is defined
by (2.18) and (2.17), (2.19) hold, then}
  $$ \sup_{|\lambda |\leq \lambda _0}|\tilde q(\lambda )-\hat q|\leq
     c|\theta |^{-1}, \qquad |\theta |\to \infty \eqno (2.21) $$
{\sl The constant $c$ in (2.21) can be chosen uniformly in
$q\in {\cal B}_C$ where ${\cal B}_C$ is defined in (1.5).}}
\par \noindent {{\bf Corollary}: {\sl The following inversion formula
holds}
  $$ \tilde q(\lambda )=-4\pi \lim_{|\theta |\to \infty }\int
 \limits_{S^2}A(\theta ',\alpha )\nu (\alpha ,\theta )\, d\alpha ,\qquad
  \theta '-\theta =\lambda ,\quad \theta',\theta \in M' \eqno (2.22) $$}
\par {Estimate (2.21) is the error estimate for the inversion formula
(2.18) for the exact data $\aaa $. Note that as the data we actually use
the coefficients $A_\ell (\alpha ),\ \ell \geq 0$. In the proof of
Theorem A the following major lemma is used:}
\par \noindent {{\bf Lemma 6}: {\sl If $\rho $ is defined by (2.12) and
$|\theta |\gg 1$, $\theta \in M$, then}
  $$ \Vert \rho \Vert _a\leq c(\Vert \rho \Vert +|\theta |^{-1})
   ,\quad \Vert\cdot\Vert_a:=\Vert\cdot\Vert_{L^2(B_a)}
    \eqno (2.23) $$ }
\par {4. We will turn now to the description of the inversion formulas
for noisy data $\adaa $ given for all $\alpha ',\alpha \in S^2$.
It will become clear later that we actually can use only a discrete
subset of noisy data $A_\delta (\alpha '_j,\alpha _p)$ for some
$\alpha '_j,\alpha _p \in S^2$. Let $[x]$ denote the integer nearest
to the real number $x$. Define
  $$ N(\delta ):=[{|\ln \delta |\over \ln |\ln \delta |}] \eqno (2.24) $$
 
  $$ \hat A_\delta (\theta ',\alpha):=\sum _{\ell =0}^{N(\delta )}
  A_{\delta \ell }(\alpha )Y_\ell (\theta '), \quad
  A_{\delta \ell }(\alpha ):=\int \limits_{S^2}\adaa
  \overline {Y_\ell (\alpha ')}\, d\alpha ' \eqno (2.25) $$
 
  $$ u_\delta (x,\alpha ):=\exp (i\alpha \cdot x)+\sum_{\ell =0}^
  {N(\delta )}A_{\delta \ell }(\alpha )Y_\ell  (\alpha ')h_\ell (r)
  \eqno (2.26) $$
 
  $$ \rho _\delta (\nu ):=\exp (-i\theta \cdot x)\int \limits_{S^2}
     u_\delta (x,\alpha )\nu (\alpha )\, d\alpha -1,\quad
     \theta \in M' \eqno (2.27) $$
 
  $$ \mu (\delta )=\exp [-\gamma N(\delta )], \quad
     \gamma :=\ln {a_1\over a}>0, \eqno (2.28) $$
 
  $$ a(\nu ):=\Vert \nu \Vert _{L^2(S^2)} \eqno (2.29) $$
{\bf Lemma 7}: {\sl One has, with $\kappa :=|$Im $\theta |$,}
  $$ \Vert \rho _\delta (\nu )\Vert \leq \Vert \rho (\nu )\Vert +
  ca(\nu )\exp (\kappa b)\mu (\delta ), \eqno (2.30) $$
 
  $$ \Vert \rho (\nu )\Vert \leq \Vert \rho _\delta (\nu )\Vert +
  ca(\nu )\exp (\kappa b)\mu (\delta ). \eqno (2.31) $$ }
\par {Consider the variational problem
  $$ |\theta |=\sup :=\Theta (\delta ),\qquad
  \Vert \rho _\delta (\nu )\Vert +a(\nu )\exp (\kappa b)\mu
  (\delta)\leq c|\theta |^{-1},\quad \theta \in M' \eqno (2.32) $$
where $c>0$ is a sufficiently large fixed constant, the supremum is taken
over $\nu\in L^2(S^2)$ and $\theta \in M',\ \delta >0$ being fixed.}
\par \noindent {{\bf Lemma 8}:
 $$ \Theta (\delta )\to \infty \hbox { as } \delta \to 0 \eqno (2.33) $$}
 See also formula (4.39) for a stronger result.
\par {Let $\nu _\delta (\alpha )\in L^2(S^2)$ be any function such that
  $$ \Vert \rho _\delta (\nu )\Vert +a(\nu _\delta )\exp (\kappa b)
  \mu _1(\delta )\leq c|\theta (\delta )|^{-1}, \quad |\theta
   (\delta )|>{\Theta (\delta )\over 2} \eqno (2.34) $$
where $\kappa =\kappa (\delta ):=|$Im $\theta (\delta )|$. One can
calculate such $\nu _\delta (\alpha )$ and $\theta (\delta )$
by solving problem (2.32). Find $\theta '(\delta )\in M'$
such that (2.19) holds with an arbitrary fixed $\lambda \in R^3$,
$|\lambda |\leq \lambda _0$. Define
  $$ \hat q_\delta :=-4\pi \int \limits_{S^2}\hat A_\delta (\theta '
     (\delta ),\alpha )\nu_\delta (\alpha )\, d\alpha \eqno (2.35) $$
Our main result is an error estimate for the inversion formula (2.35)
for noisy data. We assume that (1.3) holds and $q\in Q$ where $Q$ is
defined in Section I.}
\par \noindent {{\bf Theorem B}: {\sl If $q\in Q$ and (1.3) holds, then}
  $$ \sup_{|\lambda |\leq \lambda _0}|\tilde q(\lambda )-
  \hat q_\delta |\leq c|\theta (\delta )|^{-1}    \eqno (2.36) $$
{\sl where $\hat q_\delta $ and $\tilde q(\lambda )$ are defined in
(2.35) and (2.20),}
  $$ |\theta (\delta )|\geq {|\ln \delta |\over (\ln |\ln \delta |)^2}
  , \eqno (2.37) $$
{\sl the constant $c$ in (2.36) does not depend on $\delta $ and can
be chosen uniformly for $q\in {\cal B}_C$.}}
\par {5. If one applies a quadrature formula to (2.35) one obtains
  $$ \hat q_\delta \approx -4\pi \sum _{j=1}^n \hat A_\delta
 (\theta '(\delta ),\alpha _j)\nu _\delta (\alpha _j)c_j \eqno(2.38) $$
where $\alpha _j$ are the nodes and $c_j$ are the coefficients of a
quadrature formula. Formula (2.38) uses discrete noisy data. Using the
definition (2.25) and applying a quadrature formula again one has
  $$ \hat A_{\delta \ell }(\alpha _j )\approx \sum _{p=1}^{n_1}
     A_\delta (\alpha '_p,\alpha _j)\overline {Y_\ell (\alpha '_p)}
     c'_p\eqno (2.39) $$
Combining (2.35), (2.25), (2.38) and (2.39) one gets
  $$ \hat q\approx -4\pi \sum _{j=1}^n \nu _\delta (\alpha _j)c_j
  \sum _{\ell =0}^{N(\delta )}Y_\ell \bigl (\theta '(\delta )\bigr )
  \sum _{p=1}^{n_1}A_\delta (\alpha '_p,\alpha _j)\overline
  {Y_\ell (\alpha '_p)}c'_p \eqno (2.39') $$
This formula uses the discrete noisy data
$A_\delta (\alpha '_p,\alpha _j)$, $1\leq p\leq n_1$, $1\leq j\leq n$.
In principle one can estimate the error of the quadrature formulas
(2.37)--(2.39$'$) in terms of some bounds on the derivatives of
$\hat A_\delta (\theta '(\delta ),\alpha )$ and $\nu _\delta (\alpha )$,
but we do not go into detail.}
\par {6. Let us consider the following question. Suppose
$q_j\in Q$, $j=1,2$, $A_j(\alpha ',\alpha )$ is the scattering
amplitude corresponding to the potential $q_j$. The question is:
{\it how does one estimate $p:=q_1-q_2$ in terms of \break
$A:=A_1-A_2$.} The following lemma is useful.}
\par \noindent {{\bf Lemma 9}: {\sl One has}
  $$ -4\pi A(\alpha ',\alpha )=\int \limits_{B_a}p(x)u_1(x,\alpha )
  u_2(x,-\alpha ')\, dx \eqno (2.40) $$
{\sl where $u_j(x,\alpha )$ is the scattering solution corresponding to
the potential $q_j(x)$, $j=1,2$.}}
\par {The answer to the above question is
given in the following theorem.}
\par \noindent {{\bf Theorem C}: {\sl If}
  $$ \sup_{\alpha ',\alpha \in S^2}|A_1(\alpha ',\alpha)-
  A_2(\alpha ',\alpha)|<\delta \eqno (2.41) $$
{\sl then}
  $$ \sup_{|\lambda |\leq \lambda _0}|\tilde p(\lambda )|
  \leq c{\ln |\ln \delta |\over |\ln \delta |} \eqno (2.42) $$
{\sl where $c$=const$>0$ can be chosen uniformly for $q\in{\cal B}_C$.}}
\par \centerline {\bf III. Auxiliary Results and Proofs of Theorem A
and Lemmas 10--17}
\par {1. In this section we give some auxiliary results
which are of independent interest and also help to prove the results
formulated in Theorems A--C.}
\par {Let us start with an estimate of the quantity
$\tilde q(\lambda )$. The starting point is the well-known formula
  $$ -4\pi A(\theta ',\alpha )=\int \limits_{B_a}\exp
  (-i\theta '\cdot x)u(x,\alpha )q(x)\, dx          \eqno (3.1) $$
Multiply (3.1) by $\nu (\alpha )\in L^2(S^2)$ and integrate over $S^2$.
Write the result as
 $$ -4\pi \int \limits_{S^2}A(\theta ',\alpha )
  \nu (\alpha )\, d\alpha = \int \limits_{B_a}\exp [-i(\theta '-\theta )
  \cdot x][1+\rho (\nu )]q(x)\, dx \eqno (3.2) $$
Here $\rho (\nu )$ is defined in (2.12), $\theta '$ and $\theta $
satisfy (2.19), and (3.2) can be written as
  $$ |-4\pi \int \limits_{S^2}A(\theta ',\alpha )\nu (\alpha )\, d\alpha
-\tilde q(\lambda )|=|\int \limits_{B_a}\exp (-i\lambda \cdot x)q(x)\rho
(\nu )\,dx|\leq \Vert q\Vert _a\Vert \rho (\nu )\Vert _a \eqno (3.3) $$
where $\tilde q(\lambda )$ is defined (2.20) and
$\Vert \cdot \Vert _a=\Vert \cdot \Vert _{L^2(B_a)}$.}
\par {Let us formulate the result:}
\par \noindent {{\bf Lemma 10}: {\sl If $q\in Q_a$, $\nu \in L^2(S^2)$
and (2.19) holds then (3.3) holds.}}
\par \noindent {{\bf Proof of Theorem A}: If one uses (3.3), chooses
$\nu =\nu (\alpha ,\theta )$ such that (2.17) holds, and uses (2.23),
one obtains formula (2.21), and (2.22) follows from (2.21). Theorem A
is proved.}
\par {2. In the proof of Theorem B the following
auxiliary results are important.}
\par \noindent {{\bf Lemma 11}: {\sl One has
  $$ |\hat q_\delta -\tilde q(\lambda )|\leq 4\pi \Vert \hat
  A_\delta (\theta ',\alpha )-A(\theta ',\alpha )
  \Vert _{{}_{L^2(S^2)}} a(\nu _\delta )+\Vert \rho (\nu _\delta )
  \Vert_a \Vert q \Vert _a,                  \eqno (3.4) $$
 
  $$ \Vert \hat A_\delta (\theta ',\alpha )-A(\theta ',\alpha )\Vert _
  {L^2(S^2)}\leq c \exp (\kappa a_1)\mu (\delta )<c\exp (\kappa b)
  \mu (\delta ),                           \eqno (3.5) $$
where $\hat A_\delta (\theta ',\alpha )$ is given by (2.25),
$A(\theta ',\alpha )$ is given by (2.10), $\theta '(\delta )$ and
$\theta (\delta )$ satisfy (2.19),
  $$ \kappa :=|\hbox{Im }\theta (\delta )|,\qquad
     \mu (\delta ):=\ln [-\gamma N(\delta )],\qquad
     \gamma :=\ln {a_1\over a}>0, $$
  $$   a(\nu _\delta ):=\Vert \nu _\delta \Vert _{L^2(S^2)}, $$
$\tilde q(\lambda )$ and $\hat q_\delta $ are defined in (2.20) and
(2.35), and $N(\delta )$ is defined by (2.24).}}
\par {We collect proofs at the end of this section.}
\par {3. In the proof of Theorems B and C the following
results are used. They are of independent interest.}
\par \noindent {{\bf Lemma 12}: {\sl If $q\in Q_a$, $\theta \in M$,
$|\theta |\gg 1$ then there exists $\psi $ which satisfies the equation
  $$ \ell _q\psi =0 \hbox{ in } R^3, \quad \psi =\exp (i\theta \cdot x)
  [1+R(x,\theta )], \eqno (3.6) $$
and
  $$ \Vert R\Vert _{L^2(D)}\leq c|\theta |^{-1}, \quad
     |\theta |\to \infty \eqno (3.7) $$
where $D\subset R^3$ is an arbitrary bounded region, the constant $c$ in
(3.7) depends on diam $D$ and on
  $ \Vert q\Vert _a:=\Vert q\Vert _{L^2(B_a)} $.
  
  The following estimate holds [6] :
  $$||R||_{L^\infty (D)}<c 
\hbox{ $\displaystyle 
 (\ln |\theta|)^{ 1\over 2}\over \displaystyle |\theta|^{ 1\over 2}$}
 . \eqno (3.7')$$}}
\par {Let $\ell:=\ell _q$, $N_D(\ell ):=\{w\colon \ell w=0$ in
$D$, $w\in H^2(D)\}$, where $H^m(D)$ is the usual Sobolev space,
$D\subset R^3$ is an arbitrary bounded region with a smooth boundary.}
\par \noindent {{\bf Lemma 13}: {\sl Let
$w\in N_D(\ell )$ and $\varepsilon >0$ be an arbitrary small number.
Then there exists $\nu _\varepsilon (\alpha )$ such that
  $$ \Vert w-\int \limits_{S^2}u(x,\alpha )\nu _\varepsilon (\alpha )
  \, d\alpha \Vert _{L^2(D)}<\varepsilon \eqno (3.8) $$
Here $u(x,\alpha )$ is the scattering solution defined in
(1.1), (1.2).}}
\par \noindent {{\bf Lemma 14}: {\sl Suppose $w=\psi $ and (3.8) holds
with $w=\psi $, Im $\theta \ne 0$. Then
$$ \lim _{\varepsilon \to 0}\Vert \nu _\varepsilon (\alpha )
   \Vert =\infty,                               \eqno (3.9 ) $$
 
  $$ \Vert \nu _\varepsilon \Vert \geq c\exp (\kappa d/2),\quad \kappa :=
     |\hbox {Im }\theta |,  \quad d=\hbox{diam }D  \eqno (3.10) $$ }}
\par {Suppose that
  $$ \Vert \psi (x,\theta )-\int \limits_{S^2}u(x,\alpha )\nu
  _\varepsilon (\alpha )\, d\alpha \Vert \leq \varepsilon \eqno (3.11) $$
where the norm in (3.11) is defined in (2.14). This norm is not to be
confused with the norm in $L^2(S^2)$. It is clear from the context
what norm is meant.}
\par {Consider the problem of finding
  $$ \inf \Vert \nu _\varepsilon (\alpha )\Vert
     :=j(\varepsilon ,\theta )                 \eqno (3.12) $$
where the infimum is taken over the functions
$\nu _\varepsilon (\alpha )\in L^2(S^2)$ which satisfy (3.11).}
\par {If $\theta $, $\theta \in M$,
 is arbitrary large fixed, $\varepsilon >0$, and
  $$ n(\varepsilon )=\exp \{(\ln \ln \varepsilon ^{-1})[1+o(1)]\},\quad
     \varepsilon \to 0 \eqno (3.13) $$
then the following estimate holds.}
\par \noindent {{\bf Lemma 15}: {\sl One has
  $$ j(\varepsilon ,\theta )\leq c\exp (\kappa r)\left ({2n(\varepsilon )
  +1\over er}\right )^{n(\varepsilon )}n^2(\varepsilon ), \quad
     r>b,\quad \kappa =|\hbox{Im }\theta | \eqno (3.14) $$
where $r>b$ is arbitrary and $c$ does not depend on $\varepsilon $ and
$\theta $.}}
\par \noindent {{\bf Corollary}: {\sl Minimizing in $r>b$ one obtains
  $$ j(\varepsilon ,\theta )\leq c(2\kappa )^{n(\varepsilon )} n^2
  (\varepsilon ), \quad  \kappa =|\hbox{Im }\theta | \eqno (3.14') $$ }
\par {If $N(\kappa )$ is the asymptotic solution to the equation
  $$ \left ({eb\kappa \over N}\right )^N=c_2{\exp (-\kappa b)\over
  \kappa }, \quad \kappa \to +\infty \eqno (3.15) $$
then, with $\varepsilon (\kappa ):=\kappa\sp{-1} \exp (-\kappa b)$,
the following estimate holds.}
\par \noindent {{\bf Lemma 16}: {\sl One has
  $$ j(\varepsilon (\kappa ),\theta )\leq cN^3(\kappa )(2\kappa )^
     {2N(\kappa )+1},\quad \kappa :=|\hbox{Im }\theta |\to \infty ,
     \eqno (3.16) $$
and $N(\kappa )$ is of the order $\kappa $ as $\kappa \to \infty $:
  $$ eb\kappa <N(\kappa )<e^2b\kappa , \quad
  \kappa \to \infty                \eqno (3.16') $$
Thus
  $$ j(\varepsilon (\kappa ),\theta )\leq c\kappa ^4(2\kappa )^{2e^2b
     \kappa }\leq c\kappa ^4(2\kappa )^{20b\kappa }. \eqno (3.16'') $$}}
\par {4. Let $q\in {\cal B}_C$, where ${\cal B}_C$ is
defined in (1.5).
The scattering solution satisfies the equation
  $$ (I+T_q)u=\exp (i\alpha \cdot x),\quad T_qu:=\int \limits_{B_a}
     {\exp (i|x-y|)\over 4\pi |x-y|}q(y)\, u(y)dy \eqno (3.17) $$
Consider the operator $I+T_q$ as an operator in $L^2(B_a)$. Note that
$T_q\colon L^2(B_a)\rightarrow L^2(B_a)$ is compact.}
\par \noindent {{\bf Lemma 17}: {\sl The operator $I+T_q$ is an
isomorphism
of $L^2(B_a)$ onto $L^2(B_a)$ and
  $$ \mathop {\sup }_{q\in {\cal B}_C} \Vert (I+T_q)^{-1}\Vert \leq c
     \eqno (3.18) $$
where the constant $c$ depends on $C$ and does not depend on
$q\in {\cal B}_C$ .}}
\par {5. The following estimate holds (cf. (2.9)):
  $$ |j_\ell (r)|\leq cr^{-1/2}
  \left ({er\over 2\ell +1}\right)^
  {2\ell +1\over 2}
     {1\over (2\ell +1)^{1/2}}\, , \quad
     0\leq r\leq r_0, \quad \ell \geq 0 \eqno (3.19) $$
where $j_\ell (r)$ is the spherical Bessel function.}
\par {6. Let us prove the lemmas of this section.}
\par \noindent {{\bf Proof of Lemma 11}. One has, using (2.35),
  $$ \hat q_\delta =-4\pi \int \limits_{S^2}[\hat A_\delta
  (\theta '(\delta ),\alpha )-A(\theta '(\delta ),\alpha )]
  \nu _\delta (\alpha )\, d\alpha -4\pi \int \limits_{S^2}A(\theta '
  (\delta ),\alpha )\nu _\delta (\alpha )\, d\alpha \eqno (3.20) $$
By (3.3) one obtains from (3.20) the estimate
  $$ \eqalign {&|\hat q_\delta -\tilde q(\lambda )|\leq 4\pi \int
     \limits_{S^2}|\hat A_\delta (\theta '(\delta ),\alpha )-
 A(\theta '(\delta ),\alpha )|\, |\nu _\delta (\alpha )|\, d\alpha +\cr
     &+\Vert q\Vert _a\Vert \rho (\nu _\delta )\Vert _a\leq
     4\pi \Vert \hat A_\delta (\theta '(\delta ),\alpha )-
 A(\theta '(\delta ),\alpha )\Vert \,\Vert \nu _\delta (\alpha )\Vert +
\Vert q\Vert _a\Vert \rho (\nu _\delta )\Vert _a} \eqno (3.21) $$
This is the estimate (3.4). Let us prove (3.5)}
\par {One has
  $$ \eqalign {&\Vert \hat A_\delta (\theta '(\delta ),\alpha )-
     A(\theta '(\delta ),\alpha )\Vert \leq \Vert \sum_{\ell =0}^
     {N(\delta)} |A_{\delta \ell }(\alpha )-A_\ell (\alpha )|\,
     |Y_\ell (\theta ')|\,\Vert +\cr
     +&\Vert \sum_{\ell =N(\delta )+1}^\infty |A_\ell (\alpha )|\,
     |Y_\ell (\theta )|\, \Vert :=I_1+I_2} \eqno (3.22) $$
Using estimate (2.8) with $r=a_1>a$, and (2.9), and taking into
account that $|$Im $\theta '|=|$Im $\theta |=\kappa $ if
$\theta '-\theta =\lambda,\, \lambda \in R^3$, one obtains
  $$ \eqalign {&I_1\leq c\delta \exp (\kappa a_1)N^2(\delta )\left [
     \left ({ea_1\over 2N(\delta )+1}\right )^{2N(\delta )+1\over 2}
     {1\over \sqrt{N(\delta )}}\right ]^{-1}=\cr
     &=c\delta \exp (\kappa a_1)N^{5/2} (\delta )\left ({2N(\delta )
     +1\over ea_1}\right )^{2N(\delta )+1\over 2}\leq
     c\delta \exp (\kappa a_1)N^3(\delta ) \left (
     {N(\delta )\over ea_1/2}\right )^{N(\delta )}} \eqno (3.23) $$
Here we took into account that there are
  $$ \sum_{\ell =0}^{N(\delta )}(2\ell +1)=[N(\delta )+1]^2  $$
spherical harmonics with $0\leq \ell \leq N(\delta )$.}
\par {Furthermore, using (2.7), (2.8) and (2.9) one gets
  $$ I_2\leq \sum_{\ell =N(\delta )+1}^\infty \left (
     {ea\over 2\ell +1}\right )^{2\ell +1\over 2}{1\over 2\ell +1}
     {\exp (\kappa a_1)\over \left ({ea_1\over 2\ell +1}\right )^
     {2\ell +1\over 2}{1\over \sqrt {2\ell +1}}}\leq
     cN^{1/2}(\delta )\left ({a\over a_1}\right )^{N(\delta )}
     \exp (\kappa a_1 ) \eqno (3.24) $$
Combining (3.23) and (3.24) one has, with
$N=N(\delta ),\ a/a_1:=s,\ 0<s<1,\ ea_1/2:=t,$
  $$ I_1+I_2\leq c\exp (\kappa a_1)\left \{\delta N^3\left (
     {N\over t}\right )^N+N^{1/2}s^N\right \} \eqno (3.25) $$
\par {Consider, for a fixed small $\delta >0$, the minimization problem
$$ \delta N^3\left ({N\over t}\right )^N+N^{1/2}s^N=\inf \eqno (3.26) $$
where the infimum is taken over $N,\, N\gg 1$, that is for
sufficiently large $N$. Write (3.26) as
  $$ \delta \exp (N\ln N-N\ln t+3\ln N)+\exp (N\ln s+
      {1\over 2}\ln N)=\inf \eqno (3.27) $$
To find the infimum in $N$, let us differentiate (3.27) with respect to
$N$ and equate the resulting expression to zero. This yields
  $$ \delta \exp(N\ln N-N\ln t+3\ln N)\ln N\left [1+O\left ({1\over \ln N}
     \right )\right ]+\exp (N\ln s+{1\over 2}\ln N)[\ln s+
     {1\over 2N}]=0                \eqno (3.28) $$
Thus
  $$ \delta ^{-1}=\exp (N\ln N-N\ln t+3\ln N-N\ln s-{1\over 2}\ln N)
     {\ln N\left [1+O\left ({1\over \ln N}\right )\right ]\over
     -{1\over 2N}+\ln s^{-1}} \eqno (3.29) $$
 
  $$ \ln {1\over \delta }=N\ln N\left [1+O\left ({1\over \ln N}\right )
     \right ],\qquad N\to \infty \eqno (3.30) $$
 
  $$ \ln \ln {1\over \delta }=\ln N+\ln \ln N+O\left ({1\over \ln N}
     \right )=(\ln N)\left [1+O\left ({\ln \ln N\over \ln N}\right )
     \right ] \eqno (3.31) $$
Divide (3.30) by (3.31) to obtain
  $$ N=N(\delta ):={\ln {1\over \delta }\over \ln \ln {1\over \delta }}
     [1+o(1)], \quad \delta \to 0 \eqno (3.32) $$
Denote by $\tilde \mu (\delta )$ the infimum in (3.27) which is the
value of the left-hand side of (3.27) at $N=N(\delta )$. Then, using
(3.27) and (3.28), one gets, with $\gamma :=\ln (s^{-1})$,
  $$ \tilde \mu (\delta )=\exp \{-\gamma N(\delta )[1+o(1)]\} \left \{
     1+{\gamma -{1\over 2N(\delta )}\over [\ln N(\delta )]\left [
     1+O\left ({1\over \ln N(\delta )}\right )\right ]}\right \}\leq
     c\mu (\delta ), \eqno (3.33) $$
where $\mu (\delta ):=\exp [-\gamma N(\delta )]$.}
\par {From (3.32), (3.25) and (3.33) one obtains
  $$ I_1+I_2\leq c\exp (\kappa a_1)\mu (\delta )\leq
     c\exp (\kappa b)\mu _1(\delta ) \eqno (3.34) $$
where $\gamma _1:=\ln [a_1/(a\sqrt{2})]<\gamma ,\, b>a_1$.
 Therefore estimate
(3.5) is proved and the proof of Lemma 11 is complete. \hfill $\square$ }
\par \noindent {{\bf Proof of Lemma 12}. Substitute $\psi $ of the
form (3.6) into the equation (3.6) to get the equation for $R$:
  $$ LR:=(\nabla ^2+2i\theta \cdot \nabla )R=q(x)R+q(x)
     \hbox { in } R^3 \eqno (3.35) $$
Define the operator
  $$ L^{-1}f:=-{1\over (2\pi )^3}\int \limits_{R^3}
     {\exp (i\lambda \cdot x)\tilde f(\lambda )\over
     \lambda ^2+2\lambda \cdot \theta }\, d\lambda:=w \eqno (3.36) $$
where
  $$ \tilde f(\lambda ):=\int \limits_{R^3}
     \exp (i\lambda \cdot x)f(x) \, dx \eqno (3.37) $$
Then
  $$ Lw=f \hbox{ in } R^3 \eqno (3.38) $$
In Lemma 12 the existence of $\psi $ satisfying (3.6) and (3.7) is
claimed. Uniqueness of such a $\psi $ is not discussed. Therefore,
Lemma 12 will be proved if one demonstrates the existence of a
solution to the equation
  $$ R=L^{-1}qR+L^{-1}q \hbox { in } R^3  \eqno (3.39) $$
such that $R$ satisfies (3.7).}
\par {Suppose that the following estimate is established
  $$ \Vert L^{-1}f\Vert _{L^2(D_1)}\leq c|\theta |^{-1}
     \Vert f\Vert _{L^2(D)},\quad \theta \in M,\quad
     |\theta |\gg 1 \eqno (3.40) $$
for any $f\in L^2(R^3)$ vanishing outside of a bounded domain $D\,$,
$D\subset B_a\subset D_1\,$, $D_1$ is an arbitrary bounded domain in
$R^3$, and $c$ depends on $D$ and $D_1$ but not on $\theta $ (we will
prove estimate (3.40) later: see estimate (3.47) below).
Then equations(3.39) and (3.40) imply, with
$\Vert \cdot \Vert_a=\Vert \cdot \Vert _{L^2(B_a)}$,
  $$ \Vert R \Vert_a\leq c|\theta |^{-1}\Vert qR\Vert_a+c|\theta |^{-1}
     \Vert q\Vert_a\leq c|\theta |^{-1}\Vert q\Vert _{L^\infty (B_a)}
     \, \Vert R \Vert_a+c|\theta |^{-1}\Vert q\Vert_a \eqno (3.41) $$
Therefore, for $|\theta |\gg 1$ such that
$c|\theta |^{-1}\Vert q\Vert _{L^\infty (B_a)}<1$, one has
  $$ \Vert R \Vert_a\leq c_1 |\theta |^{-1}\Vert q\Vert_a\eqno (3.42) $$
This implies (3.7) under the additional assumption
$q\in L^\infty (B_a)$. 

To complete the proof of Lemma 12 under this
additional assumption it is sufficient to prove (3.40). This
estimate is an immediate consequence of the following result, which
can be found in [0, vol. II, pp. 17, 34]: Let $P(\partial )$
be a partial differential expression with constant coefficients,
$\partial :=-i$ grad, $D\subset R^n,\, n\geq 2$, $D$
is a bounded domain,
 $G_0(x)$
is a regular fundamental solution: $P(\partial )G_0=\delta (x-y)$,
$\eta (x)\in C_0^\infty (R^n)$, $\eta =1$ in a neighborhood of the set
$\{x-y\}$, $x,y\in D_1$, and $G_1(x-y)=\eta (x-y)G_0(x-y)$. Then
  $$ \int \limits_D G_1(x-y)f(y)\, dy=\int \limits_D G_0(x-y)f(y)\, dy,
     \quad D\subset D_1, \quad x\in D_1\eqno (3.43) $$
and
  $$ \sup_{\lambda \in R^n}{\cal P}(\lambda )
     |\tilde G_1(\lambda )|<\infty \eqno (3.44) $$ }
where
$$ {\cal P}(\lambda ):=(\sum_{|j|\geq 0} |P^{(j)}
   (\lambda )|^2)^{1/2}\eqno (3.45) $$
and $j$ is a multiindex. If
$P(\partial )=L:=\nabla ^2+2i\theta \cdot \nabla$, then
$L(\lambda )=\lambda ^2+2\lambda \cdot \theta $ and
  $$ \eqalign {&{\cal L}(\lambda )=(\sum_{|j|\geq 0} \vert (\lambda^2
     +2\lambda \cdot \theta ) ^{(j)}\vert ^2)^{1/2}= \cr
    =&\{|\lambda ^2+2\lambda \cdot \theta |^2+4|\lambda +\theta |^2+12\}
     ^{1/2}\geq 2|\hbox {Im }\theta |} \eqno (3.46) $$
Therefore
  $$ \eqalign {&\Vert L^{-1}f\Vert_{L^2(D_1)}^2=\int \limits_{D_1}
     \vert \int \limits_D G_1(x-y)f(y)\,dy\vert ^2\, dx\leq
   {1\over (2\pi )^3}\int \limits_{R^3}d\lambda |\tilde G_1(\lambda )|^2
  |\tilde f(\lambda )|^2\leq \cr   &\leq c\max _{\lambda \in R^3}
  [{\cal L}^2(\lambda )|\tilde G_1(\lambda )|^2]\Vert f
  \Vert _{L^2(D)}^2\cdot {1\over
  {\min_{\lambda \in R^3}{\cal L}^2(\lambda )}}\leq c
  |\theta |^{-2}\Vert f \Vert _{L^2(D)}^2} \eqno (3.47) $$
Estimate (3.47) is identical with (3.40). Therefore Lemma 12 is proved
under the additional assumption $q\in L^\infty (B_a)$. Without this
assumption estimate (3.7) is proved in [6] and in [1]. The argument in
[6] is more complicated. It uses  estimate (3.7'):
  $$ \Vert L^{-1}f\Vert _{L^\infty (D_1)}\leq c|\theta |^{-1/2}(\ln
    |\theta |)^{1/2}\,\Vert f\Vert _{L^2(D)},\quad |\theta |\gg 1, \quad
     \theta \in M  \eqno (3.48) $$
In place of the estimate (3.41) one uses the estimate
  $$ \Vert R\Vert _{L^\infty (D_1)}\leq c(|\theta |^{-1}\ln |\theta |)^{1/2}
  \, \Vert q\Vert _{L^2(B_a)}\Vert R\Vert _{L^\infty (D_1)}+c(|\theta |
     ^{-1}\ln |\theta |)^{1/2}\, \Vert q\Vert_{L^2(B_a)} \eqno (3.49) $$
which implies
  $$ \Vert R\Vert _{L^\infty (D_1)}\leq c(|\theta |^{-1}\ln |\theta |)^{1/2},
  \quad q\in Q_a,\quad \theta \in M,\quad |\theta |\gg 1 \eqno (3.50) $$
Using (3.50) and the first inequality (3.41) one obtains
  $$ \Vert R\Vert_a \leq c|\theta |^{-1}\Vert R\Vert _{L^\infty (D_1)}
    \Vert q\Vert_a+c|\theta |^{-1}\Vert q\Vert 
    \leq c_1|\theta |^{-1}                  \eqno (3.51) $$
Lemma 12 is proved. \hfill $\square$ }
\par \noindent {{\bf Proof of Lemma 13}. Suppose that $w\not \equiv 0$
and (3.8) is false. Then $w$ is orthogonal in $L^2(S^2)$ to all
functions of the form
  $$ \int \limits_{S^2}u(x,\alpha )\nu (\alpha )\, d\alpha , $$
that is
  $$ 0=\int \limits_D\overline{w(x)}\int \limits_{S^2}u(x,\alpha )
     \nu (\alpha ) \, d\alpha \, dx \, \hbox { for all }
      \nu \in L^2(S^2) \eqno (3.52) $$
Thus
  $$ 0=\int \limits_D\overline{w(x)}u(x,\alpha )\, dx \,
  \hbox { for all } \alpha \in S^2          \eqno (3.53) $$
It is proved in [3, p. 46] that
  $$ G(x,y)={\exp (i|y|)\over 4\pi |y|}u(x,\alpha )+o(|y|^{-1})
  \hbox{ as } |y|\to \infty ,\quad {y\over |y|}=-\alpha \eqno (3.54) $$
where $G(x,y)$ is the resolvent kernel for
$\ell_q\colon \ell_qG=-\delta (x-y)$ in $R^3$, $G$ satisfies the
radiation condition, $\ell_q$ is defined in (1.1). Therefore (3.53)
implies
  $$ 0=\int \limits_D\overline{w(x)}G(x,y)\, dx:=h(y) \,
  \hbox { for all } y\in D'                 \eqno (3.55) $$
The function $h(y)$ satisfies the equations
  $$ \ell_qh=-\bar w(x) \hbox { in } R^3 \eqno (3.56) $$
Since $h\in H_{\rm loc}^2$, it follows from (3.55) that
  $$ h=h_N=0 \hbox { on } \partial D \eqno (3.57) $$
Multiply (3.56) by $w(x)$, integrate over $D$ and then by parts, using
(3.57) and the equation $\ell_qw=0$ in $D$, to get
  $$ 0=\int \limits_D |w|^2\, dx \eqno (3.58) $$
Therefore $w=0$. This contradiction proves Lemma 13. \hfill $\square$ }
\par \noindent {{\bf Proof of Lemma 14}. Suppose (3.9) is false. Then
  $$ \Vert \nu_\varepsilon (\alpha )\Vert \leq c \hbox { for all }
\varepsilon \in (0,\varepsilon_0), \quad \varepsilon_0>0 \eqno (3.59) $$
Choose a weakly convergent in $L^2(S^2)$ subsequence and denote it
$\nu_\varepsilon (\alpha )$ again, $\varepsilon \to 0$. Pass to the limit
$\varepsilon\to 0$ in (3.8) with $w=\psi (x,\theta )$ and this subsequence
$\nu_\varepsilon (\alpha )$, $\nu_\varepsilon (\alpha )\to \nu (\alpha )$
weakly in $L^2(S^2)$. The result is
  $$ \psi (x,\theta )=\int \limits_{S^2}u(x,\alpha )\nu (\alpha )\,
    d\alpha \hbox { in }D,\quad \hbox {Im }\theta \ne 0 \eqno (3.60) $$
Since both sides in (3.60) solve the elliptic equation
    $$ \ell_q\psi =0 \hbox { in } R^3 \eqno (*) $$
they are identical in $R^3$ by the unique continuation property for the
solution to ($\ast $). This is a contradiction since $\psi (x,\theta )$
grows exponentially as $|x|\to \infty $ in some directions, while the
integral
   $$ |\int \limits_{S^2}u(x,\alpha )\nu (\alpha )\, d\alpha |\leq
      c\Vert \nu \Vert $$
is bounded. Here we used the well known estimate
  $$ \mathop {\sup }_{x\in R^3,\alpha \in S^2}|u(x,\alpha )|\leq c
     \eqno (3.61) $$
Estimate (3.9) is proved. In order to prove (3.10) one deduces from
(3.8) with $w=\psi (x,\theta )$ that
$$ \Vert \int \limits_{S^2}u(x,\alpha )\nu_\varepsilon (\alpha ,\theta )
  \, d\alpha \Vert _{L^2(D)}\geq \Vert \psi \Vert_{L^2(D)}-\varepsilon
\geq c\exp (\kappa d/2),\quad \kappa =|\hbox{Im }\theta | \eqno (3.62) $$
where $d=$diam $D$. Assume that for some $\varepsilon >0$ the estimate
(3.10) is false. Then there is a sequence $\theta _n\in M$,
$|\theta _n|\to \infty $, such that
  $$ \Vert \nu _\varepsilon \Vert \exp (-\kappa_nd/2)\to 0
  \eqno (3.63) $$
This contradicts inequality (3.62). Indeed, from (3.62) it follows that
  $$ 0<c\leq \exp (-\kappa_nd/2)\,\Vert \int \limits_{S^2}u(x,\alpha )
     \nu_\varepsilon (\alpha ,\theta_n)\, d\alpha )\,\Vert_{L^2(D)}\leq
     c_1\exp (-\kappa_n d/2)\Vert \nu_\varepsilon (\alpha ,\theta_n)
     \Vert_{L^2(S^2)}  \eqno (3.64) $$
where $c_1>0$ does not depend on $\varepsilon$ or $\theta_n$, it depends
on $\max_{x\in R^3,\alpha \in S^2}|u(x,\alpha )|$ (see (3.61)) and on
\break (meas $D)^{1/2}$. Since (3.64) contradicts (3.63), Lemma 14 is
proved. \hfill $\square$ }
\par {We will use Lemma 17 in the proof of Lemmas 15 and 16. Therefore
let us first prove Lemma 17.}
\par \noindent {{\bf Proof of Lemma 17}: First, let us prove that
$N(I+T_q)=\{0\}$, where $N(A)$ is the null space of a linear operator
$A$. It is easy to see that, for $q\in Q_a$, the operator $T_q$ in
(3.17) is a linear compact operator on $L^2(B_a)$. Therefore
$(I+T_q)^{-1}$ exists and is bounded by the Fredholm alternative if
$N(I+T_q)=\{0\}$. Suppose $w+T_qw=0$ in $B_a$. Define $w(x)$ to be
$-T_qw$ for any $x\in R^3$. Then $w+T_qw=0$ in $R^3$. Therefore $w$
solves the problem:
  $$ [\nabla^2+1-q(x)]w=0 \hbox { in } R^3,\qquad |x|\left ({\partial w
 \over \partial |x|}-iw\right )\to 0,\quad |x|\to \infty \eqno (3.65) $$
It is well known [19] that (3.65) implies $w=0$. Therefore
$N\{I+T_q\}=\{0\}$ and the operator $(I+T_q)^{-1}$ exists and is
bounded in $L^2(B_a)$. Let us prove (3.18). Assume that (3.18) is false.
Then there is a sequence $q_n\in {\cal B}_C$ such that
  $$ \Vert (I+T_n)^{-1}\Vert \geq n, \quad T_n:=T_{q_n}. \eqno (\ast) $$
We prove that this is impossible. Indeed, since ${\cal B}_C$ is a
compact set in $L^2(B_a)$, one can select a convergent in $L^2(B_a)$
subsequence which we denote again by $q_n$, $\Vert q_n-q\Vert_a \to 0$
as $n\to \infty $. One can check that
$$ \Vert T_{q_{_1}}-T_{q_{_2}}\Vert \leq c\Vert q_1-q_2\Vert_a
       \eqno (3.66) $$
Indeed
  $$ \Vert T_{q_{_1}}f-T_{q_{_2}}f\Vert^2 \leq {1\over (4\pi)^2}\left (
  \int \limits_{B_a}\, dx
  \int \limits_{B_a} {|q_1-q_2|^2\over |x-y|^2}\, dy
  \right )\Vert f\Vert^2\leq c\Vert q_1-q_2\Vert_a^2 \Vert f\Vert^2 $$
from which (3.66) follows with $c=c(a)$. Therefore, if $q_n\to q$ in
$L^2(B_a)$ then $\Vert T_n-T\Vert \to 0$ as $n\to \infty $. One has,
using bounded invertibility of $I+T$,
  $$ \Vert (I+T_n)^{-1}\Vert =\Vert (I+T+T_n-T)^{-1}\Vert =\Vert
     (I+T)^{-1}[I+(T_n-T)(I+T)^{-1}]^{-1}\Vert \leq c  \eqno (3.67) $$
Here the inequality holds for all sufficiently large $n$, for example,
for such $n$ that $\Vert T_n-T\Vert \, \Vert (I+T)^{-1}\Vert <1$. Since
(3.67) contradicts $(\ast )$, Lemma 17 is proved. \hfill $\square$ }
\par \noindent {{\bf Proof of Lemma 15}: Assume that $\theta \in M$,
$|\theta |\gg 1$, is fixed. First, note that one can find
$\nu (\alpha )$ such that $\Vert \nu \Vert $ is arbitrary large but
 $$ \Vert \int \limits_{S^2}u(x,\alpha )\nu (\alpha )\, d\alpha
 \Vert_{L^2(D)}    \leq \delta , \eqno (\ast ) $$
where $\delta >0$ is an arbitrary small number. Indeed, $u=Bu_0$,
$u_0:=\exp (i\alpha \cdot x)$ and $B=(I+T)^{-1}$ is a bounded linear
operator in $L^2(D)$, $\Vert B\Vert +\Vert B^{-1}\Vert \leq c$ (see Lemma
17). Since $\Vert h\Vert =\Vert B^{-1}Bh\Vert \leq \Vert B^{-1}\Vert $
$\Vert Bh\Vert $, one concludes from $(\ast )$ that
  $$ c\,\Vert \int \limits_{S^2}\exp (i\alpha \cdot x)\nu (\alpha )\,
     d\alpha \Vert_{L^2(D)}\leq \delta , \eqno (\ast \ast ) $$
where $c$ does not depend on $\delta $. Conversely, $(\ast \ast )$
implies $(\ast )$ (with $c_1\delta $ in place of $\delta $)
since $B$ is a bounded operator. Take
$\nu_\ell (\alpha )=c_\ell Y_\ell (\alpha )$,
$\Vert \nu \Vert =|c_\ell |$. Take $|c_\ell |\to \infty $ as
$\ell \to \infty $ so that $|c_\ell j_\ell (d)|\to 0$ as
$\ell \to \infty $ where $d=$diam $D$. Such a choice of $c_\ell $ is
possible since $|j_\ell (b)|\to 0$ as $\ell \to \infty $ (see (3.19)).
So, $\Vert \nu_\ell (\alpha )\Vert \to \infty $, and
  $$ \Vert \int \limits_{S^2}u(x,\alpha )\nu_\ell (\alpha )\, d\alpha
  \Vert_{L^2(D)}\leq c\,\Vert \int \limits_{S^2}\exp (i\alpha \cdot x)
     c_\ell Y_\ell (\alpha )\, d\alpha \Vert_{L^2(D)}=c'|c_\ell |\,
    \Vert j_\ell (r)\Vert_{L^2(D)}\to 0 \hbox { as }\ell \to \infty $$
The claim is proved. That is why we are looking for
$\nu_\varepsilon (\alpha )$ which satisfies (3.11) and has minimal norm.}
\par {Let us note that $\psi =B\varphi $, where $\varphi $ solves the
equation
  $$ \nabla^2\varphi +\varphi =0 \hbox { in } R^3 \eqno (3.68) $$
Indeed, since $\varphi =B^{-1}\psi =(I+T)\psi $, one has
$(\nabla^2+1)\varphi =(\nabla^2+1)(I+T)\psi=q\psi -q\psi =0$ as claimed.
Inequality (3.11) implies, as in the passage from ($\ast $) to
($\ast \ast $), that
  $$ \Vert \varphi-\int \limits_{S^2}\exp (i\alpha \cdot x)
     \nu_\varepsilon (\alpha )\, d\alpha \Vert \leq c\varepsilon :=
     \varepsilon_1, \quad b>a_1>a\sqrt{2}, \eqno (3.69) $$
where $c$ does not depend on $\varepsilon $ and $\theta $, the
norm $\Vert \cdot \Vert $ in this argument is equivalent to the norm
$\Vert \cdot \Vert_{L^2(B_b)}$, and
  $$ \varphi =\psi +T\psi =\psi +\int \limits_{B_a}{\exp (i|x-y|)\over
     4\pi |x-y|}q(y)\psi (y,\theta )\, dy \eqno (3.70) $$
Since $b>a$ and $|\theta |\gg 1$, $\Vert \varphi \Vert_b$ is of order
of $\Vert \psi \Vert_b$ which, in turn, is of order of
$\Vert \exp (i\theta \cdot x)\Vert_b$ as $|\theta |\to \infty $.
Also, $\Vert \varphi \Vert_b$, $\Vert \psi \Vert_b$ and
$\Vert \exp (i\theta \cdot x)\Vert_b$ are of order
$\Vert \exp (i\theta \cdot x)\Vert $ as $|\theta |\to \infty $,
$\theta \in M$, where $\Vert \cdot \Vert $ is defined in (2.14). Since
$\varphi $ solves (3.68) one can write
  $$ \varphi (x)=\sum_{\ell=0}^\infty a_\ell j_\ell (r)
    Y_\ell (\alpha '),\qquad  r=|x|,\quad \alpha '=x/r \eqno (3.71) $$
Also
  $$ \exp (i\alpha \cdot x)=\sum_{\ell =0}^\infty 4\pi i^\ell
     j_\ell (r)Y_\ell (\alpha ')\overline{Y_\ell (\alpha )}, \qquad
     r=|x|,\quad \alpha '=x/r  \eqno (3.72) $$
Thus (3.69) can be written as
  $$ \sum_{\ell =0}^\infty |a_\ell -4\pi i^\ell \nu_{\varepsilon\ell}|^2
  b_\ell ^2\leq \varepsilon_1 ^2,\qquad b_\ell ^2:=\int \limits_{a_1}^b
      r^2|j_\ell (r)|^2\, dr \qquad \nu_{\varepsilon \ell }
     :=(\nu_\varepsilon ,Y_\ell)_{L^2(S^2)}  \eqno (3.73) $$
One can write an equation similar to (3.17) for $\psi $:
  $$ (I+\Gamma )\psi =\exp (i\theta \cdot x),\quad \Gamma\psi :=\Gamma_q
\psi :=
     \int \limits_{B_a}G(x,y,\theta )q\psi \, dy \eqno (3.74) $$
Here $G$ solves the equation
  $$ (\nabla^2+1)G=-\delta (x), \quad  G=\exp (i\theta \cdot x)g
     \eqno (3.75) $$
 
  $$ Lg:=(\nabla^2+2i\theta \cdot \nabla )g=-\delta (x), \qquad
     g={1\over (2\pi )^3}\int \limits_{R^3}{\exp (i\lambda \cdot x)\over
     \lambda^2+2\lambda \cdot \theta }\, d\lambda \eqno (3.76) $$
The operator $(I+\Gamma )^{-1}$ exists and is bounded in $L^2(B_a)$ if
$|\theta|\gg 1$.
To prove this, it is sufficient, as in the proof of Lemma 17, to prove
that $N(I+\Gamma )=\{0\}$. Suppose $w+\Gamma w=0$ in $B_a$. Then $w$
solves equation (3.65) and satisfies the following condition at
infinity: $w=\exp (i\theta \cdot x)v$, where
  $$ v=L^{-1}qv \hbox { in } R^3 \eqno (3.77) $$
and $L^{-1}$ is defined in (3.36). The relation (3.77) plays the role
of the condition at infinity. Using estimate (3.40) one derives from
(3.77) restricted to $B_a$ that $v=0$ provided that $|\theta |\gg 1$,
namely, $|\theta |$ is so large that
       $ \Vert L^{-1}q\Vert_{L^2(B_a)\rightarrow L^2(B_a)}<1$.
Thus $v=0$, $w=0$, and the operator $(I+\Gamma )^{-1}$ is bounded in
$L^2(B_a)$. Since $\Gamma $ is a compact linear
integral operator in $L^2(B_a)$
and $I+\Gamma $ is injective,
one can write $(I+\Gamma )^{-1}=I+\Gamma_1$ where $\Gamma_1$ is a
compact linear integral operator in $L^2(B_a)$. Therefore
  $$ \psi =\exp (i\theta \cdot x)+\Gamma_1\exp (i\theta \cdot x)=
     \exp (i\theta \cdot x)+\int \limits_{B_a}\Gamma_1(x,y)
      \exp (i\theta \cdot y)\, dy \eqno (3.78) $$
and from (3.70) one obtains
  $$ \varphi =\exp (i\theta \cdot x)+\Gamma_1\exp (i\theta \cdot x)+T
 \exp (i\theta \cdot x)+T\Gamma_1\exp (i\theta \cdot x) \eqno (3.79) $$
It follows from (3.79) that $\exp (i\theta \cdot x)$ is the main term
of $\varphi $ as $|\theta |\to \infty $ in the  region $|x|>a$.
Since $\varphi =\psi +T\psi $ and $|\psi |\leq c\exp (\kappa r)$
 one has
$|\varphi |\leq c\exp (\kappa r),\ r>a$. Thus
$$\| \varphi\| ^2_{L^2(S^2)} =\sum \limits _{\ell =0} ^{\infty }
|a_\ell |^2j_\ell^2(r)<c\exp (2\kappa r), \quad r>a \eqno (3.80) $$
and
$$ |a_\ell |<c\exp (\kappa r)|j_\ell (r)|^{-1}, \quad r>a \eqno (3.81) $$
Define
$$ \nu_{\varepsilon \ell }:=(4\pi i^\ell )^{-1}a_\ell
\quad \hbox {for } l\leq N, \quad
   \nu_{\varepsilon \ell }:=0 \hbox { for } \ell >N \eqno (3.82) $$
Then, using (3.80) and (2.9), one gets
  $$ \eqalign {&\Vert \nu_\varepsilon \Vert^2=\sum_{\ell =0}^\infty
|\nu_{\varepsilon \ell }|^2=\sum_{\ell =0}^N|\nu_{\varepsilon \ell }|^2
     \leq {\exp (2\kappa r)\over 4\pi }\sum_{\ell =0}^N{2\ell +1\over
     |j_\ell (r)|^2}\leq \cr  \leq &c\exp (2\kappa r)\sum_{\ell =0}^N
     \left ({2\ell +1\over er}\right )^{2\ell +1}(2\ell +1)^2\leq c\exp
     (2\kappa r)\left ({2N+1\over er}\right )^{2N+1}N^3} \eqno (3.83) $$
Let us fix $r>b$ and estimate $b_\ell $ in
(3.73) using (3.19):
  $$ b_\ell ^2\leq \int \limits_0^b r\left ({er\over 2\ell +1}\right )
     ^{2\ell +1}{1\over 2\ell +1}\, dr \leq c{e^{2\ell +1}b^{2\ell +3}
     \over (2\ell +1)^{2\ell +2}(2\ell +3)}\leq c{(eb)^{2\ell }\over
     (2\ell )^{2\ell +3}}\leq c\left ({eb\over 2\ell +1}\right )^
     {2\ell +1}{1\over \ell^2} \eqno (3.84) $$
From (3.81), (3.82) and (3.84) one concludes that (3.73) holds if
  $$ \eqalign {&c\sum_{\ell =N+1}^\infty \exp (2\kappa r)\left [\left (
     {er\over 2\ell +1}\right )^{2\ell +1}{1\over 2\ell +1}\right ]^{-1}
     \left ({eb\over 2\ell +1}\right )^{2\ell +1}{1\over \ell^2}\leq \cr
     &\leq c\exp (2\kappa r)\sum_{\ell =N+1}^\infty \left ({b\over r}
     \right )^ {2\ell +1}\leq c\exp (2\kappa r)\left ({b\over r}
     \right )^{2N}\leq \varepsilon_1^2, \quad r>b} \eqno (3.85) $$
Write (3.85) as
   $$ \exp (\kappa r)\left ({b\over r}\right )^N\leq \varepsilon_2,
      \quad \varepsilon_2=\varepsilon_1/\sqrt{c}:=c_2\varepsilon,
      \quad  N\gg 1     \eqno (3.86) $$
Since $r>b$ is arbitrary, one can minimize the left-hand side of
(3.86) in $r$ and get
   $$ \inf_{r>b}\left [\exp (\kappa r)\left ({b\over r}\right )^N
      \right ]=\exp (N) \left ({b\kappa \over N}\right )^N, $$
the infimum being attained at $r=N/\kappa $. Consider the equality in
(3.86) with this $r$:
   $$ \left ({eb\kappa \over N}\right )^N=\varepsilon_2, \quad
      \varepsilon_2\to 0   \eqno (3.87) $$ }
\par {Let us solve (3.87) for $N$ asymptotically
as $\varepsilon_2\to 0$, for a fixed $\kappa $. Write (3.87) as
  $$ \ln \ln \varepsilon_2^{-1}=\ln N+\ln \ln N-{\ln (eb\kappa )\over
     \ln N}+o\left ({1\over \ln N}\right )=(\ln N)\left [1+O\left (
     {\ln \ln N\over \ln N}\right )\right ] \eqno (3.88) $$
Therefore
  $$ \ln N=(\ln \ln \varepsilon_2^{-1}) [1+o(1)],\quad
     \varepsilon_2\to 0    \eqno (3.89) $$
Thus
  $$ N=\exp \{(\ln \ln \varepsilon_2^{-1})[1+o(1)]\}, \qquad
    \varepsilon_2=c_2\varepsilon,\quad \varepsilon\to 0 \eqno (3.90) $$
With $N=N(\varepsilon )$ given by (3.90) formula (3.83) yields an upper
estimate for $\Vert \nu_\varepsilon \Vert $ and therefore for
$j(\varepsilon ,\theta )$ for arbitrary large fixed $\theta \in M$ and
$\varepsilon \to 0$:
  $$ j(\varepsilon ,\theta )\leq c\exp (\kappa r)\left ({2N+1\over er}
     \right )^N N^2  \eqno (3.91) $$
where $N$ is given by (3.90). Minimize the right-hand size of (3.91) in
$r$, $r>b$. As above, the minimum is attained at $r=N/\kappa $ and
(3.91) with this $r$ yields
  $$ j(\varepsilon ,\theta )\leq c\left ({2N+1\over N}\kappa \right )^N
     N^2\leq c(2\kappa )^N N^2 \eqno (3.92) $$
Since $N$ in (3.90) can be written as $n(\varepsilon )$ in (3.13),
Lemma 15 is proved. \hfill $\square$ }
\par \noindent {{\bf Proof of Lemma 16}: If
$\varepsilon =\exp (-\kappa b)/\kappa $, $\kappa =|$Im $\theta |$,
then (3.87) takes the form
  $$ \left ({eb\kappa \over N}\right )^N=c_2{\exp (-\kappa b)\over
     \kappa } \eqno (3.93) $$
Let $N=N(\kappa )$ solve (3.93) asymptotically as $\kappa \to +\infty $.
Denote $\nu_\varepsilon (\alpha ,\theta )$ corresponding to
$\varepsilon =\varepsilon (\kappa ):=\kappa ^{-1}\exp (-\kappa b)$ by
$\nu_\kappa $. The Fourier coefficients of $\nu_\kappa $ are given by
(3.82) with $N=N(\kappa )$. From (3.83) one derives
  $$ j(\varepsilon (\kappa ),\theta )\leq c{\exp (2\kappa r)\over
     r^{2N(\kappa )+1}}\left ({2N(\kappa )+1\over e}\right )
     ^{2N(\kappa )+1} N^3(\kappa ), \quad r>b  \eqno (3.94) $$
Minimizing in $r>b$ the right-hand side of (3.94) one obtains, at
$r=(2N(\kappa )+1)/(2\kappa )$,
  $$ \eqalign {&j(\varepsilon ,\theta )\leq c{\exp [2N(\kappa )+1]\over
     \left ({2N(\kappa )+1\over 2\kappa }\right )^{2N(\kappa )+1}}\left
     ({2N(\kappa )+1\over e}\right )^{2N(\kappa )+1} N^3(\kappa )=\cr
=&c\exp [2N(\kappa )+1]\left ({2\kappa \over e}\right )^{2N(\kappa )+1}
  N^3(\kappa )=cN^3(\kappa )(2\kappa )^{2N(\kappa )+1}} \eqno (3.95) $$
Estimate (3.16) is proved. To prove (3.16$'$), note that if
$N(\kappa )<eb\kappa $ or $N(\kappa )>e^2b\kappa $ then (3.15) cannot
hold as $\kappa \to \infty $. Indeed, as $\kappa \to \infty $ then
$(eb\kappa /N)^N>1>c_2\exp (-\kappa b)/\kappa $ for
$N(\kappa )<eb\kappa $ and
  $$ \left ({eb\kappa \over N}\right )^N<\left ({1\over e}\right )^N
     < \exp (-e^2b\kappa )<c_2\kappa^{-1}\exp (-\kappa b)
     \hbox { for } N(\kappa )>e^2b\kappa . $$
Lemma 16 is proved. \hfill $\square$ }
\par \centerline {\bf IV. Proofs of Lemmas 1--9}
\par \noindent {{\bf Proof of Lemma 1}: One has, using (3.72) and the
orthonormality of the spherical harmonics
  $$ \eqalign {&\vert A_{\ell m}(\alpha )|={1\over 4\pi }|\int
     \limits_{B_a} q(x)u(x,\alpha )\int \limits_{S^2}\exp (-i\beta
     \cdot x)\overline{Y_{\ell m}(\beta )}\, d\beta \, dx\vert \leq \cr
     &\leq c\int \limits_{B_a}|q(x)j_{\ell }(r)Y_{\ell m}(\alpha ')|
     \, dx, \qquad r=|x|, \quad \alpha '=x/r}  \eqno (4.1)  $$
and $c$ is the constant from (3.61). From (4.1) using (3.19) one obtains
  $$ |A_{\ell m}(\alpha )|\leq c\,\Vert q\Vert_a\left (\int \limits_0^a
  r^2|j_\ell (r)|^2\, dr\right )^{1/2}\leq c_1a
  \left ({ea\over 2\ell +1}
  \right )^{(2\ell +1)/2}{1\over 2\ell +1} \eqno (4.2) $$
Lemma 1 is proved. \hfill $\square $}
\par \noindent {{\bf Proof of Lemma 2}: Using the formula
  $$ \int \limits_{S^2}\exp (i\theta \cdot \alpha r)Y_\ell (\alpha )\,
  d\alpha =4\pi i^\ell j_\ell (r)Y_\ell (\theta ),\quad \theta \in M
  \eqno (4.3) $$
where $r>0$ is arbitrary, one applies the Cauchy inequality to the
integral in (4.3) and obtains
  $$ |Y_\ell (\theta )|\leq {\exp (\kappa r)\over 4\pi |j_\ell (r)|}
  \Vert Y_\ell (\alpha )\Vert \biggl (\int \limits_{S^2}d\alpha \biggr )
  ^{1/2}={\exp (\kappa r)\over \sqrt{4\pi }|j_\ell (r)|} \eqno (4.4) $$
Lemma 2 is proved. \hfill $\square $}
\par \noindent {{\bf Proof of Lemma 3}: The function $w$, defined by the
right-hand side of (2.15), solves the Helmholtz equation
  $$ (\nabla^2+1)w=0 \hbox { for } r>a \eqno (4.5) $$
and has the asymptotics
  $$ w=\exp (i\alpha \cdot x)+\aaa r^{-1}\exp (ir)+o(r^{-1}), \qquad
     r\to \infty ,\quad \alpha '=x/r \eqno (4.6) $$
Therefore the function $v:=w-u(x,\alpha )$, where $u(x,\alpha )$ is the
scattering solution, solves equation (4.5) in the region $r>a$ and
  $$ v=o(r^{-1}),\qquad r\to \infty \eqno (4.6') $$
By the Rellich-type uniqueness lemma (see e.g. [3, p. 24]) one has
$v=0$ for $r>a$. Lemma 3 is proved. \hfill $\square $}
\par \noindent {{\bf Remark}: Lemma 3 is well known.}
\par \noindent {{\bf Proof of Lemma 4}: One has, with
$\varphi :=\int \limits_{S^2}u(x,\alpha )\nu (\alpha )\, d\alpha ,$
  $$ \eqalign {&\Vert \rho (\nu )\Vert :=\Vert \exp (-i\theta \cdot x)
     \int \limits_{S^2}u(x,\alpha )\nu (\alpha )\, d\alpha -1\Vert =\cr
     =&\Vert \exp (-i\theta \cdot x)[\varphi -\exp (i\theta \cdot x)-\exp
     (i\theta \cdot x)R(x,\theta )+\exp (i\theta \cdot x)R(x,\theta )]
     \Vert \leq \cr      &\Vert \exp (-i\theta \cdot x)[\varphi -\psi ]
     \Vert +\Vert R(x,\theta )\Vert }   \eqno (4.7)  $$
By Lemma 13, with $w=\psi $ and $\psi $ as in Lemma 12, one can find
$\nu =\nu (\alpha ,\theta )$ such that
  $$ \Vert \varphi -\psi \Vert \leq {\exp (-\kappa b)\over \kappa }\, ,
     \quad \kappa :=|\hbox {Im } \theta | \eqno (4.8) $$
With this choice of $\nu $ one has
  $$ \Vert \exp (-i\theta \cdot x)[\varphi -\psi ]\Vert \leq \exp
  (\kappa b)\Vert \varphi -\psi \Vert \leq \kappa^{-1}  \eqno (4.9) $$
Since $\theta \in M$ one has $|\theta |/\kappa \to \sqrt{2}$ as
$\theta \to \infty $. From this, (4.7), (4.8) and (4.9) the estimate
(2.16) follows. Lemma 4 is proved. \hfill $\square $}
\par \noindent {{\bf Proof of Lemma 5}: Let us choose the 
coordinate system so that $\lambda=te_3$, where $t>0$ and 
$e_j,\,j=1,2,3,$ is an orthonormal basis of $R^3$. Let
$\theta ':={1 \over 2 }te_3+v,\, \theta:=-{ 1\over 2 }te_3 +v$,
where $v\cdot v=1-t^2/4, \, v_3=0, \, |v| \to \infty, \, v\in C^2$.
Clearly there are infinitely many such $v$. 
If $t\neq 1$, then one can choose $\theta '$ and $\theta$
in $M'$ in the above proof.
Lemma 5 is proved. \hfill $\square $} 

The reader can also consult [4] for another proof.

\par \noindent {The proof of Lemma 6  is given
in Section V.}        \par {Let us prove the remaining lemmas.}
\par \noindent {{\bf Proof of Lemma 7}: One has
  $$ \eqalign {&\Vert\rho_\delta (\nu )
\Vert :=\Vert \exp (-i\theta \cdot x)\int
\limits_{S^2}u_\delta (x,\alpha )\nu(\alpha )\, d\alpha -1\Vert \leq \cr
     &\leq \Vert \exp (-i\theta \cdot x)\int \limits_{S^2}u(x,\alpha )
     \nu(\alpha )\, d\alpha -1 \Vert +\Vert \exp (-i\theta \cdot x)
     \int \limits_{S^2}[u_\delta (x,\alpha )-u(x,\alpha )
     ]\nu (\alpha ) \, d\alpha \Vert \leq \cr         &\leq \Vert \rho
     (\nu )\Vert + c\exp (\kappa b)a(\nu ) \sup_{\alpha \in S^2}
     \Vert u_\delta (x,\alpha )-u(x,\alpha )\Vert } \eqno (4.10) $$ }
\par {Let us prove that
  $$ \sup_{\alpha \in S^2} \Vert u_\delta (x,\alpha )-u(x,\alpha )\Vert
  \leq c\mu(\delta )                  \eqno (4.11) $$
where $\mu(\delta )$ is defined in (2.28) and $N(\delta )$ is defined
in (2.24). One has, with $N=N(\delta )$, $r=|x|$ and
$\alpha '=x/r$,
  $$ \Vert u_\delta (x,\alpha )-u(x,\alpha )\Vert \leq \Vert
     \sum_{\ell= 0}^N[A_{\delta \ell }(\alpha )-A_\ell (\alpha )]Y_\ell
     (\alpha ')h_\ell (r)\Vert +\Vert \sum_{\ell =N+1}^\infty A_\ell
   (\alpha )Y_\ell (\alpha ')h_\ell (r)\Vert :=I_1+I_2 \eqno (4.12) $$
Using Parseval's equality and the assumption (1.3) one gets
  $$ I_1^2=\max_{\alpha \in S^2}\sum_{\ell =0}^N |A_{\delta \ell }
     (\alpha )-A_\ell (\alpha )|^2\int \limits_{a_1}^b r^2|h_\ell (r)|^2
     \, dr \leq 4\pi \delta^2 \max_{0\leq \ell \leq N}\int
     \limits_{a_1}^b r^2|h_\ell (r)|^2\, dr.  \eqno (4.13) $$
Furthermore
  $$ I_2^2\leq \max_{\alpha \in S^2}\sum_{\ell =N+1}^\infty
     |A_\ell (\alpha )|^2 \int \limits_{a_1}^b r^2|h_\ell (r)|^2\, dr
     \eqno (4.14) $$
Define
 $$ H_\ell :=\int \limits_{a_1}^b r^2|h_\ell (r)|^2\, dr \eqno (4.15) $$
One can prove (see [5]) that
  $$ H_\ell \leq c\left ({2\ell +1\over ea_2}\right )^{2\ell }
     \ell^{1/2}, \quad a_2={a_1\over \sqrt{2}}    \eqno (4.16) $$
In fact, a stronger estimate holds: 
$$H_\ell \leq ca_1^2
[(2\ell +1)\ell]^{-1} \left ({2\ell +1\over ea_1}
\right )^{2\ell+1}.\eqno (4.16')
$$
This follows from the known asymptotics of Hankel's functions:
$h_\ell (r) \sim -i (r\ell )^{-0.5}\left ({2\ell +1\over er}
\right )^{\ell + 0.5 }$ as $\ell \to \infty$.

From (4.13)--(4.16') and (2.7) one gets
  $$ I_1+I_2\leq c\left [\delta \left ({2N+1\over ea_1}\right )^{N+1}+
     \left ({a\over a_1}\right )^N\right ] \eqno (4.17) $$
Minimizing with
respect to $N$ the expression in brackets in (4.17) for a small
fixed $\delta >0$ one obtains that the minimum is $c\mu_1(\delta )$
with $\mu_1(\delta )$ defined in (2.28) and $N=N(\delta )$ defined in
(2.24) (cf. [5]). Inequality (2.30) is proved. Inequality (2.31) can be
proved similarly. Lemma 7 is proved. \hfill $\square $}
\par \noindent {{\bf Proof of Lemma 8}: Use (2.30) to conclude that
(2.32) holds if
  $$ \Vert \rho (\nu )\Vert +a(\nu )\exp (\kappa b)\mu (\delta )
     \leq c|\theta |^{-1 } \eqno (4.18) $$
Using Lemma 13 with $\varepsilon =|\theta |^{-1}\exp (\kappa b)$,
$\kappa =\vert $Im $\theta \vert $ and $D=B_b$, choose
$\nu =\nu (\alpha ,\theta )$ such that
  $$ \Vert \varphi -\psi \Vert \leq |\theta |^{-1}\exp (-\kappa b)
     \eqno (4.19) $$
where
  $$ \varphi :=\int \limits_{S^2} u(x,\alpha )\nu
  (\alpha ,\theta )\, d\alpha                \eqno (4.20) $$
Then, using (3.6), (3.7) and (4.19), one gets
  $$ \eqalign {&\Vert \rho (\nu )\Vert =\Vert \exp (-i\theta \cdot x)
     \varphi -1\Vert \leq \Vert \exp (-i\theta \cdot x)[\varphi -\psi ]
     \Vert + \Vert R \Vert \leq \cr
     &\leq \exp (\kappa b)\Vert \varphi -\psi \Vert +c|\theta |^{-1}
      \leq |\theta |^{-1}(c+1)}  \eqno (4.21) $$
Choose $\theta =\theta (\delta )$ such that
  $$ |\theta |a\bigl (\nu (\alpha ,\theta )\bigr )\exp (\kappa b)
     =1/\mu(\delta )  \eqno (4.22) $$
Note that $\mu^{-1}(\delta )\to \infty $ as $\delta \to 0$.
We claim that
  $$ a\bigl (\nu (\alpha ,\theta )\bigr )\to \infty \hbox { as }
     |\theta |\to \infty , \quad \theta \in M    \eqno (4.23) $$
From (4.23) it follows that equation (4.22) has a solution
$\theta (\delta )$ such that
  $$ |\theta (\delta )|\to \infty \hbox { as } \delta \to 0
     \eqno (4.24) $$
To finish the proof, let us check that (4.23) holds. Assume the
contrary, that is
  $$ a\bigl (\nu (\alpha ,\theta )\bigr ) \leq c'\eqno (4.25) $$
where $c'$ does not depend on $\theta $. Using (4.25) choose a
weakly convergent in $L^2(S^2)$ subsequence which is denoted by
$\nu (\alpha ,\theta_n)$:=$\nu_n$,
  $$ \nu (\alpha ,\theta_n)\to \nu (\alpha ) \hbox { weakly in }
      L^2(S^2) \hbox{ as } |\theta_n |\to \infty \eqno (4.26) $$
Pass to the limit in (4.19) to get
  $$ \Vert \psi (x,\theta_n)-\varphi (x,\theta_n )\Vert \to 0,\quad
     n\to \infty  \eqno (4.27) $$
where
  $$ \varphi (x,\theta_n):=\int \limits_{S^2}u(x,\alpha )
  \nu (\alpha ,\theta_n )\, d\alpha     \eqno (4.28) $$
This is a contradiction since
$\Vert \varphi \Vert \leq c\Vert \nu_n\Vert (4\pi )^{1/2}\leq
cc'(4\pi )^{1/2}\leq c_1$,
while $\Vert \psi (x,\theta_n)\Vert \to \infty $ as $n\to \infty $.
In this estimate $c$ is the constant from (3.61). Lemma 8 is proved.
\hfill $\square $}
\par \noindent {{\bf Remark}: The rate of growth of
$a\bigl (\nu (\alpha ,\theta )\bigr )$ as $|\theta |\to \infty $,
$\theta \in M$, can be estimated from below by formula (3.10) and from
above, for some choice of $\nu (\alpha ,\theta )$, by
(3.16)--(3.16$''$). This allows one to get the following lower estimate
of $\Theta (\delta )$ which gives a refinement of (2.33):
  $$ \Theta (\delta )\geq |\theta (\delta )|\geq c\, { |\ln \delta |
     \over ( \ln |\ln \delta |)^2}  \eqno (4.29) $$
To prove (4.29) let us solve (4.22), asymptotically as $\delta \to 0$,
for $|\theta (\delta )|$ using the upper estimate (3.16$''$) for
$a(\nu ,\theta )$. Equation (4.22) becomes
  $$ |\theta |\exp (\kappa b)\kappa^4(2\kappa )^{20b\kappa }=
     1/\mu(\delta )  \eqno (4.30) $$
Note that in order to obtain a lower estimate for $|\theta (\delta )|$
one has to use an upper estimate for $a(\nu )$ in (4.22). Since
$|\theta |/\kappa \to \sqrt{2}$ as $|\theta |\to \infty $,
$\theta \in M$, one can write (4.30) as
  $$ \exp [\kappa b+5\ln \kappa +20b\kappa \ln \kappa +20b\kappa \ln 2]
     =\exp [\gamma N(\delta )]  \eqno (4.31) $$
where $N(\delta )$ is given in (2.24). Note that the main term of the
asymptotic solution to (4.31) will be the same for the equation
(4.31) with $c\exp [\gamma N(\delta )]$ in place of
$\exp [\gamma N(\delta )]$, $c=$const. From (4.31) one derives
  $$ 20b\kappa (\ln \kappa) \left [1+\left ({1\over \ln \kappa }\right )
     \right ]=\gamma N(\delta ) \eqno (4.32) $$
Thus
  $$ \kappa \ln \kappa =\gamma_1 N(\delta )[1+o(1)], \qquad
     \delta \to 0, \quad \gamma_1:={\gamma \over 20b} \eqno (4.33) $$
Taking $\ln $ of (4.33) and denoting $\gamma_1N(\delta ):=n$ yields
  $$ \ln \kappa +\ln \ln \kappa =\ln n+o(1)  \eqno (4.34) $$
Let us look for the asymptotic solution to (4.34) of the form
  $$ \kappa ={n\over \ln n}(1+t),\quad t=o(1) \hbox { as } n\to \infty
     \eqno (4.35) $$
Substitute (4.35) into (4.34) to get
  $$ \ln n-\ln \ln n+t+O(t^2)+\ln \left \{(\ln n)\left [
     1-{\ln \ln n+t+O(t^2)\over \ln n}\right ]\right \}=\ln n+o(1) $$
Thus
  $$ t+O(t^2)-{\ln \ln n+t+O(t^2)\over \ln n}=o(1)  \eqno (4.36) $$
Equation (4.36) implies that $t=o(1)$ as $n\to \infty $. Therefore
formula (4.35) gives an asymptotic solution to equation (4.33) as
$\delta \to 0$. Let us write this solution for references in terms of
$|\theta (\delta )|=\sqrt{2}\kappa (\delta )[1+o(1)]$ as $\delta \to 0$
  $$ |\theta (\delta )|={\sqrt{2}\gamma_1N(\delta )\over \ln [\gamma_1N
     (\delta )] }[1+o(1)],\qquad \delta \to 0,\quad \gamma_1={\gamma
     \over 20b}  \eqno (4.37) $$
where $N(\delta )=|\ln \delta |/\ln |\ln \delta |$. Therefore
  $$ |\theta (\delta )|={\sqrt{2}\gamma_1|\ln \delta |\over
     \ln |\ln \delta | [\ln \gamma_1+\ln |\ln \delta |-
     \ln \ln |\ln \delta |]}[1+o(1)]={\sqrt{2}\gamma_1|\ln \delta |\over
     (\ln |\ln \delta |)^2}[1+o(1)], \quad \delta \to 0 \eqno (4.38) $$
Let us formulate the result:}
\par \noindent {{\bf Lemma 8$'$}: {\sl One has
  $$ \Theta (\delta )\geq |\theta (\delta )|={\sqrt{2}\gamma_1
     |\ln \delta |\over (\ln |\ln \delta |)^2}[1+o(1)] \hbox { as }
     \delta \to 0  \eqno (4.39) $$
where
  $$ \gamma_1={\ln [a_1/a]\over 20b}>0  \eqno (4.40) $$ }}
\par \noindent {{\bf Proof of Lemma 9}: The starting point is the
standard resolvent identity
  $$ G_1(x,y)-G_2(x,y)=-\int \limits_{B_a}G_1(x,z)p(z)G_2(z,y)\, dz
     \eqno (4.41) $$
where $p:=q_1-q_2$ and $G_j$ is the resolvent kernel of the operator
$\ell_j:=\nabla^2+1-q_j(x)$. Let
$|y|\to \infty $, $y/|y|=\beta $ in (4.41). Using (3.54) one gets
  $$ u_1(x,-\beta )-u_2(x,-\beta )=-\int \limits_{B_a}G_1(x,z)p(z)
     u_2(z,-\beta )\, dz \eqno (4.42) $$
Let $|x|\to \infty $, $x/|x|=-\alpha $ in (4.42) and use (1.2) and
(3.54) again to get
  $$ A_1(-\alpha ,-\beta )-A_2(-\alpha ,-\beta )={-1\over 4\pi }\int
  \limits_{B_a} p(z)u_1(z,\alpha )u_2(z,-\beta )\, dz \eqno (4.43) $$
By the well known reciprocity property
$A_j(-\alpha ,-\beta )=A_j(\beta ,\alpha )$, so (4.43) is identical
with (2.40). The original proof of Lemma 9, given in [9], was a little
longer. \hfill $\square $}
\par \centerline {\bf V. Proof of Lemma 6}
\par {This proof requires some preparation. Consider the equation
  $$ L\rho :=(\nabla^2+2i\theta \cdot \nabla )\rho=v \hbox { in } R^3,
     \qquad \rho \in C_0^2(B_r),\quad \theta \in M \eqno (5.1) $$
Let
  $$ D_j:=-i{\partial \over \partial x_j},\qquad D=-i\nabla ,\qquad
     \partial_j={\partial \over \partial x_j}       \eqno (5.2) $$
 
  $$ P(\xi ):=\xi^2+2\beta \cdot \xi ,\qquad \beta :=h\theta \quad
     h:=|\theta |^{-1}                    \eqno (5.3) $$
 
  $$ \beta \cdot \beta =h^2,\qquad |\beta |=1      \eqno (5.4) $$
 
  $$ N:=\{\xi \colon P(\xi )=0,\quad \xi \in R^3\} \eqno (5.5) $$
 
  $$ N_h:=\{\xi \colon \xi \in R^3, \quad \hbox {dist }(\xi ,N)\leq h\}
     \eqno (5.6) $$
 
  $$ N'_h:=R^3\setminus N_h                     \eqno (5.7) $$
Let $\beta =m+i\mu,\, m,\mu \in R^3$. Then
  $$ N=\{\xi \colon |\xi+m|=|m|,\quad \mu \cdot \xi =0,
  \quad \xi \in R^3\}     \eqno (5.8) $$
Thus $N$ is a circle. Let $P(\xi )=P_1(\xi )+iP_2(\xi )$, where
$P_1(\xi ):=$Re $P(\xi )$. Note that
  $$ dP_1(\xi )\neq 0 \hbox { for } \xi \in N         \eqno (5.9) $$
where $dP_1$ is the differential of $P_1$. Equation (5.1) can be
written as
  $$ P(hD)\rho :=[(hD)^2+2\beta \cdot hD]\rho =-h^2v   \eqno (5.10) $$
Define
  $$ F_hu:=\hat u(\xi ):=(2\pi h)^{-3/2}\int \limits_{R^3} u(x)
     \exp (-i\xi \cdot xh^{-1})\, dx      \eqno (5.11) $$
 
  $$ u(x)=(2\pi h)^{-3/2}\int \limits_{R^3}\hat u(\xi )\exp
     (i\xi\cdot xh^{-1})\, d\xi  \eqno (5.12) $$
 
  $$ ih\partial_{\xi_j}\hat u(\xi )=\widehat {x_ju} \eqno (5.13) $$
 
  $$ F_h\{-ih\partial_ju(x)\}=\xi_j\hat u(\xi ) \eqno (5.14) $$ }
\par {Let us denote in this section
  $$ \Vert \rho \Vert :=\Vert \rho \Vert_{L^2(R^3)} \eqno (5.15) $$
 
  $$ \Vert \rho \Vert_{b_1,b_2}:=\Vert \rho \Vert_{L^2(B_{b_2}\setminus
     B_{b_1})},\quad 0<b_1<b_2    \eqno (5.16) $$
 
  $$ \Vert g\langle hD\rangle \rho \Vert :=\Vert g(\sqrt {1+\xi^2})
     \hat \rho (\xi )\Vert   \eqno (5.17) $$
First, we need}
\par \noindent {{\bf Lemma 18}: {\sl Any solution $\rho \in C_0^2(B_r)$
to (5.10) satisfies the inequality}
  $$ h\Vert \langle hD\rangle^2\rho \Vert \leq c\Vert P(hD)\rho \Vert
     \quad \forall h\in (0,h_0)  \eqno (5.18) $$
{\sl where $h_0>0$ is a fixed sufficiently small number.}}
\par \noindent {{\bf Proof}: Inequality (5.18) can be written as
  $$ h\Vert (1+\xi^2)\hat \rho \Vert \leq c\Vert P(\xi )\hat \rho \Vert
     \eqno (5.19) $$
where the definition (5.17) and Parseval's equality are used.}
\par {Let $\xi \in N'_h$. Then
  $$ h(1+\xi^2)\leq c|P(\xi )|                \eqno (5.20) $$
so that
  $$ h^2\int \limits_{N'_h}(1+\xi^2)^2|\hat \rho |^2\, d\xi \leq c^2\int
     \limits_{N'_h}|P(\xi )|^2|\hat \rho (\xi)|^2\, d\xi \leq c^2
     \int \limits_{R^3}|P(\xi )\hat \rho(\xi )|^2\, d\xi =c^2\int
     \limits_{R^3}|P(hD)\rho |^2 \, dx       \eqno (5.21) $$
If $\xi \in N_h$, then one introduces local coordinates in which the
set $N$ has the equations
  $$ t=0,\quad \xi_1=0,\quad t=P_1(\xi ),     \eqno (5.22) $$
the $\xi_1$-axis being chosen along the vector $\mu $ defined by the
equation $\beta =m+i\mu $. These local coordinates can be introduced
because of the condition (5.9). Define
  $$ f:=P_1(\xi )\hat \rho (\xi )       \eqno (5.23) $$
Then $f=0$ at $t=0$ and $f\in C^\infty (R^3)$. Let us use the inequality
  $$ \int \limits_{-h}^h t^{-2}|f(t)|^2\, dt \leq 4 \int \limits_{-h}^h
     |f'(t)|^2\, dt   \eqno (5.24) $$
proved in Lemma 19 below. This inequality holds for
  $$ f\in C^1(-h,h),\qquad h={\rm const}>0,\quad f(0)=0 \eqno (5.25) $$
Applying (5.24) to (5.23) and integrating (5.24) in the remaining
variables $\xi $, one gets
  $$ \int \limits_{N_h}|\hat \rho (\xi )|^2\,d\xi \leq c\int
     \limits_{N_h} |\nabla_\xi [P_1(\xi )\hat \rho (\xi )]|^2\, d\xi \leq
     c\int \limits_{R^3}|\nabla_\xi [P_1(\xi )\hat \rho (\xi )]|^2\,
     d\xi     \eqno (5.26) $$
Note that the set $N_h$ is bounded. Therefore
  $$ h^2\int \limits_{N_h}(1+\xi^2)^2|\hat \rho (\xi )|^2\, d\xi \leq
     ch^2\int \limits_{N_h}|\hat \rho (\xi )|^2\, d\xi \eqno (5.27) $$
Using Parseval's equality, the assumption supp $\rho \subset B_r$,
and the S.Bernstein's inequality for derivatives of entire
functions of exponential type,
one gets
  $$ h^2\int \limits_{R^3}|\nabla_\xi [P_1(\xi )\hat \rho (\xi )]|^2\,
     d\xi =r^2\int \limits_{R^3}|P_1(\xi)\hat\rho |^2dx\leq
     r^2\int \limits_{R^3}|P(\xi)\hat\rho |^2\, dx=r^2\int \limits_{R^3}
     |P(hD)\rho |^2\, dx    \eqno (5.28) $$
From (5.26)--(5.28) one obtains
  $$ h^2\int \limits_{N_h}(1+\xi^2)^2|\hat \rho (\xi )|^2\, d\xi \leq
     cr^2\int \limits_{R^3}|P(hD)\rho |^2\, dx   \eqno (5.29) $$
From (5.21) and (5.29) one gets
  $$ h^2\int \limits_{N_h}(1+\xi^2)^2|\hat \rho (\xi )|^2\, d\xi \leq
     c^2\int \limits_{R^3}|P(hD)\rho |^2\, dx  \eqno (5.30) $$
This inequality implies (5.18). Lemma 18 is proved.}
\par \noindent {{\bf Lemma 19}: {\sl Under the assumptions (5.25)
inequality (5.24) holds.}}
\par \noindent {{\bf Proof}: Inequality (5.24) is similar to the
well known Hardy's inequality in which the integration is taken over
$(0,\infty )$. To prove (5.24) one starts with an obvious inequality
in which $\lambda $ is an arbitrary real number:
  $$ 0\leq \int \limits_{-h}^h|f'-\lambda t^{-1}f(t)|^2\, dt:=
     A\lambda^2-\lambda B+C        \eqno (5.31) $$
where
  $$ A:=\int \limits_{-h}^h t^{-2}|f|^2\, dt,\qquad B:=\int
     \limits_{-h}^h t^{-1}{d\over dt}|f(t)|^2\, dt   \eqno (5.32) $$
and
  $$ C:=\int \limits_{-h}^h |f'|^2\, dt              \eqno (5.33) $$
One has
  $$ B=t^{-1}|f(t)|^2\vert_{-h}^h +\int \limits_{-h}^h t^{-2}|f|^2\, dt
     ={|f(h)|^2+|f(-h)|^2\over h}+A\geq A    \eqno (5.34) $$
From (5.31) it follows that
  $$ B^2\leq 4AC              \eqno (5.35) $$
By (5.34) one has $B\geq A$ and (5.35) implies
  $$ A\leq 4C                 \eqno (5.36) $$
This is inequality (5.24). Lemma 19 is proved.   \hfill $\square $}
\par {We need one more lemma:}
\par \noindent {{\bf Lemma 20}: {\sl Let}
  $$  P(hD)\rho =0 \hbox { in } A_1      \eqno (5.37) $$
{\sl where $A_1$ is a bounded region with a smooth boundary. Let
$A\subset A_1$, $\eta (x)\in C_0^\infty (A_1)$,
$0\leq \eta (x)\leq 1$, $\eta (x)=1$ in  $A$, where $A$ is a strictly
inner subdomain of $A_1$. Then}
  $$ h\Vert D\rho \Vert_A\leq c\Vert \rho \Vert_{A_1}  \eqno (5.38) $$
{\sl where $\Vert \rho \Vert_A:=\Vert \rho \Vert_{L^2(A)}$.}
\par \noindent {{\bf Proof}: Multiply (5.37) by $\eta \bar \rho $, take
the real part and integrate by parts to get
  $$ \eqalign {&h\int \limits_{A_1}\eta |\nabla \rho |^2\, dx=
     -{h\over 2}\int \limits_{A_1}(\bar \rho \nabla \rho +\rho \nabla
    \bar \rho ) \nabla \eta \, dx+2\hbox {Re }
\left( i\beta_j\int \limits_{A_1}
     \rho_j \bar \rho \eta \, dx\right) =\cr     &={h\over 2}\int \limits_{A_1}
     |\rho |^2\nabla^2\eta \, dx +2\hbox {Re }\left( i\beta_j\int \limits_{A_1}
     \rho_j \bar \rho \eta \, dx\right) }   \eqno (5.39) $$
where the summation over the repeated indices is understood. Using the
inequalities
  $$ |2\rho_j\rho |\leq {h\over 2}|\rho_j|^2+2h^{-1}|\rho |^2
     \eqno (5.40) $$
 
  $$ |\nabla^2 \eta |\leq c, \quad |\beta_j|\leq 1   \eqno (5.41) $$
one obtains from (5.39) the following inequality
  $$ h\int \limits_{A_1}\eta |\nabla \rho |^2\, dx\leq ch\int
     \limits_{A_1}|\rho |^2\, dx+{h\over 2}\int \limits_{A_1}\eta
     |\rho_j|^2\, dx+2h^{-1}\int \limits_{A_1}\eta |\rho |^2\, dx
     \eqno (5.42) $$
It follows from (5.42) that
  $$ h^2\int \limits_A |\nabla \rho |^2\leq c_1h^2\int \limits_{A_1}
     |\rho |^2\, dx+c_2\int \limits_{A_1}\eta |\rho |^2\, dx \leq
     c_3\int \limits_{A_1}|\rho |^2\, dx      \eqno (5.43) $$
This inequality implies (5.38). Lemma 20 is proved. \hfill $\square $}
\par \noindent {We are now ready to prove Lemma 6.}
\par \noindent {{\bf Proof of Lemma 6}: The equation for the function
$\rho $ defined in (2.12) is
  $$ L\rho =q\rho +q \hbox { in } R^3       \eqno (5.44) $$
where $L$ is defined in (5.1)  Write (5.44) as
  $$ P(hD)\rho =-h^2(q\rho +q), \quad h:=|\theta |^{-1} \eqno (5.45) $$
Let $\eta \in C_0^\infty (B_b)$, $0\leq \eta (x)\leq 1$,
$\eta (x)=1$ in $B_{a_1}$. Clearly
  $$ P(\eta \rho )=(P\eta -\eta P)\rho -h^2\eta (q\rho +q)
     \eqno (5.46) $$
Applying (5.18) to (5.46) yields
  $$ h\Vert \langle hD\rangle^2(\rho \eta) \Vert \leq c\Vert (P\eta -\eta P)
     \rho \Vert +ch^2\Vert q\Vert_{L^{^\infty }(B_a)}
     \Vert \rho \Vert_a + ch^2\Vert q \Vert_a      \eqno (5.47) $$
Since $\eta (x)=1$ in $B_a$, one obtains
  $$ h\Vert \rho \Vert_a\leq h\Vert \langle hD\rangle^2(\eta \rho)\Vert
     \leq ch^2\Vert \rho \Vert_a+ch^2+c\Vert (P\eta -\eta P)\rho \Vert
     \eqno (5.48) $$
Thus
  $$ \Vert \rho \Vert_a \leq ch+ch^{-1}\Vert (P\eta -\eta P)\rho \Vert
     \eqno (5.49) $$
Using the equation
  $$ D\eta =0 \hbox { in } B_{a_1}        \eqno (5.50) $$
one obtains
  $$ \Vert (P\eta -\eta P)\rho \Vert =\Vert \rho (hD)^2\eta +2h^2D\eta
     \cdot D\rho +2h\beta \rho \cdot D\eta \Vert \leq c(h^2+h)\Vert
     \rho \Vert_{a_1,b}+ch^2\Vert D\rho \Vert_{a_1,b} \eqno (5.51) $$
The function $\rho $ solves equation (5.44)
and $q(x)=0$ in $B_b\setminus B_{a_1}$. Therefore Lemma 20 is
applicable and the estimate (5.38) yields
  $$ h\Vert D\rho \Vert_{a_1,b}\leq c\Vert \rho
     \Vert_{a_1-\varepsilon ,b+\varepsilon }   \eqno (5.52) $$
where $\varepsilon >0$ is an arbitrary small number.}
\par {From (5.49), (5.51)and (5.52) one obtains
  $$ \Vert \rho \Vert_a\leq ch+c\Vert \rho
     \Vert_{a_1-\varepsilon ,b+\varepsilon }   \eqno (5.53) $$
Since $\varepsilon >0$ is arbitrarily small, one chooses
$\varepsilon $ so small that $a<a_1-\varepsilon $, where $a_1>a$, and
(5.53) implies (2.23). Lemma 6 is proved. \hfill $\square $}
\par \noindent{The arguments in this section are close to those in [8].}
\par \centerline {\bf VI. Proof of Theorems B and C}
\par {We have already proved Theorem A in Section III.1.}
\par \noindent {{\bf Proof of Theorem B}: From (2.35) with
$\theta =\theta (\delta )$ and $\theta '=\theta '(\delta )$, where
$\theta (\delta ),\theta '(\delta )$ and $\nu_\delta (\alpha )$ are
defined in Lemma 8, one has
  $$ \hat q_\delta =-4\pi \int \limits_{S^2}[\hat A_\delta
     (\theta',\alpha)-A(\theta',\alpha)]\nu_\delta (\alpha )\, d\alpha-
     4\pi\int \limits_{S^2}A(\theta',\alpha)\nu_\delta (\alpha )\,
     d\alpha :=I_1+I_2                \eqno (6.1) $$
The last term in (6.1) is transformed as in (3.2) and (3.3) and one
obtains the estimate
  $$ \vert -4\pi\int \limits_{S^2}A(\theta',\alpha)\nu_\delta (\alpha )
     \, d\alpha -\tilde q(\lambda )\vert \leq \Vert q \Vert_a \Vert
     \rho (\nu_\delta )\Vert_a\leq c|\theta |^{-1}     \eqno (6.2) $$
where formula (2.32) was used.}
\par {The first term is estimated by (3.5) and (2.32):
  $$ |I_1|\leq ca(\nu_\delta )\exp (\kappa b)\mu(\delta )\leq
     c|\theta |^{-1}                     \eqno (6.3) $$
From (6.2) and (6.3) the estimate (2.36) follows.}
\par {Let us prove (2.37). To this end one has to estimate the function
$\Theta (\delta )$, defined in  (2.32), from below. Such an
estimate is given in (4.39) and yields (2.37). Finally, Lemma 17 shows
that the constants $c$ in our estimates can be chosen uniformly for
$q\in {\cal B}_C$. Theorem B is proved.   \hfill $\square $}
\par \noindent {{\bf Proof of Theorem C}: The starting point is formula
(2.40). Multiply (2.40) by $\nu_1(\alpha ,\theta )$ and
$\nu_2(-\alpha ',\theta ')$ and integrate over $S^2\times S^2$ to get
  $$ -4\pi \int\limits_{S^2}\int\limits_{S^2}\aaa \nu_1(\alpha ,\theta )
    \nu_2(-\alpha ',\theta ')\, d\alpha \, d\alpha '=\int \limits_{B_a}
     p(x)\varphi_1(x,\theta )\varphi_2(x,\theta ')\, dx  \eqno (6.4) $$
Here $\theta ,\theta '\in M,\quad |\theta |\gg 1$,
$\theta '+\theta =\lambda $, $\lambda \in R^3$,
  $$ \varphi_j(x,\theta_j):=\int \limits_{S^2}u_j(x,\alpha )
     \nu_j(\alpha ,\theta_j)\, d\alpha,\quad j=1,2, \qquad
     \theta_1:=\theta ,\quad \theta_2:=\theta '   \eqno (6.5) $$
and $\nu_j=\nu_j(\alpha ,\theta_j)$ are chosen so that
  $$ \Vert \rho (\nu_j)\Vert \leq c|\theta |^{-1}, \quad
     \rho (\nu_j)=\exp (-i\theta_j \cdot x)\varphi_j-1 \eqno (6.6) $$
where $|\theta_1 |\,|\theta_2 |^{-1}\to 1$, as $|\theta_1|\to \infty $,
and the norm is defined in (2.14). Then, using (6.6) and (2.23), one
obtains
  $$ \int \limits_{B_a}p(x)\varphi_1\varphi_2\, dx=\int \limits_{B_a}
     p(x)\exp (i\lambda \cdot x)(1+\rho_1)(1+\rho_2)\, dx=
     \tilde p(\lambda )+O(|\theta |)^{-1})    \eqno (6.7) $$
Note that for $|\lambda |\leq \lambda_0$ one can choose a constant $c$
independent of $\lambda $, such that
  $$ O(|\theta |^{-1})\leq c|\theta |^{-1}, \qquad {|\theta_1|\over
     |\theta_2|}\to 1 \hbox { as } |\theta_1|\to \infty \eqno (6.8) $$
The left-hand side of (6.4) can be estimated by the Cauchy inequality.
Using (2.41) one gets
  $$ \vert -4\pi \int \limits_{S^2}\int \limits_{S^2}\aaa \nu_1\nu_2\,
     d\alpha \, d\alpha '\vert \leq 16\pi^2\delta a(\nu_1)a(\nu_2)
     \eqno (6.9) $$
From (6.4) and (6.7)--(6.9) one obtains
  $$ |\tilde p(\lambda )|\leq c[\delta a(\nu_1)a(\nu_2)+|\theta |^{-1}]
     \eqno (6.10) $$}
\par {Let us estimate $a(\nu_j)$. If one takes
$\varepsilon =\kappa^{-1}\exp (-\kappa b)$ in (3.11), then estimates
(6.6) hold (see formulas (4.7)--(4.9)). Therefore, by (3.16$''$),
  $$ a(\nu_j)\leq c\kappa^4(2\kappa )^{20b\kappa }, \quad
     \kappa =|\hbox {Im }\theta_1|=|\hbox {Im }\theta_2| \eqno (6.11) $$
Thus (6.10) and (6.11) yield
  $$ |\tilde p(\lambda )|\leq c[\delta \kappa^8(2\kappa )^{40b\kappa }+
     \kappa^{-1}]       \eqno (6.12) $$
For a small fixed $\delta >0$ find the infimum of the right-hand side
of (6.12)
  $$ \inf_{\kappa >0} [\delta \kappa^8(2\kappa )^{40b\kappa }+
     \kappa^{-1}]:=\eta (\delta )   \eqno (6.13) $$
One has
  $$ \delta h(\kappa )+\kappa^{-1}:=\delta \exp (40b\kappa \ln \kappa +
     40b\kappa \ln 2+8\ln \kappa )+\kappa^{-1}=\inf    \eqno (6.14) $$
Taking the derivative with respect to $\kappa $ yields, at the point of
minimum,
  $$ \delta h(\kappa )[40b\ln \kappa +O(1)]=\kappa^{-2}, \quad
     \kappa \gg 1          \eqno (6.15) $$
Let us solve (6.15) asymptotically, as $\delta \to 0$, for
$\kappa =\kappa (\delta )$. This allows us to estimate $\eta (\delta )$
as $\delta \to 0$. Taking $\ln $ of (6.15) yields
  $$ \ln {1\over \delta }=2\ln \kappa +40b\kappa \ln \kappa +40b\kappa
     \ln 2+8\ln \kappa +\ln [40b\ln \kappa +O(1)]=
     40b\kappa (\ln \kappa)
     [1+o(1)],\quad \kappa \to 0       \eqno (6.16) $$
This is an equation similar to (4.33). Using the argument given for
the proof of Lemma 8$'$, one obtains the asymptotic solution (6.16).
The role of $N(\delta )$ is played by $\ln (1/\delta )$, and the role
of $\gamma_2$ is played by $1/(40b)$. Thus, (4.35) yields
  $$ \kappa (\delta )={E(\delta )\over \ln E(\delta )}[1+o(1)],
  \hbox { where }    E(\delta ):={1\over 40b}\ln {1\over \delta},
     \quad \delta \to 0                   \eqno (6.17) $$
Equation (6.17) can be written as
  $$ \kappa (\delta )={1\over 40b}{|\ln \delta |\over \ln |\ln \delta |}
     [1+o(1)],     \quad \delta \to 0       \eqno (6.18). $$
From (6.13)--(6.15) and (6.18) one obtains
  $$ \eta (\delta )={1\over \kappa }[1+o(1)]=40b{\ln |\ln \delta |\over
     |\ln \delta |}[1+o(1)] \hbox { as } \delta \to 0 \eqno (6.19) $$
This and (6.12) yield (2.42). The constant $c$ by Lemma 17 can be
chosen uniformly for $q\in {\cal B}_C$. Theorem C is proved.
\hfill $\square $}
\par \noindent {Theorem C is a refinement and an improvement over the
result in [14] (see also [31]).}
\par \centerline {\bf VII. Summary of the Numerical Procedure,
Additional Results and Remarks}
\par {In this last section we discuss the following items: a) stability
of the recovery of $q(x)$ given $\tilde q(\lambda )$ with some error,
b) discussion of the steps in a possible numerical implementation of
the methods for solving the ISP developed in this paper, and c) an
alternative to (2.32) optimization problem which leads algorithmically
to a stable approximation of $\tilde q(\lambda )$ from the knowledge
of noisy data.}
\par {1. Suppose $\tilde q_\delta (\lambda )$ is known such that
  $$ \sup_{|\lambda |\leq \lambda_0}|\tilde q_\delta (\lambda ) -
     \tilde q(\lambda )|\leq \delta       \eqno (7.1) $$
where $\lambda_0>0$ is a large number, $\tilde q(\lambda )$ is defined
in (2.20), and a priori it is assumed that $q(x)\in Q$ and
  $$ |\tilde q(\lambda )|\leq c_1(1+\lambda^2)^{-\tilde d}, \qquad
   \tilde d>{3\over 2}, \quad \lambda \in R^3     \eqno (7.2) $$
Given $\tilde q_\delta (\lambda )$ for $|\lambda |\leq \lambda_0$,
 and the
numbers $c_1,\tilde d$ and $\delta $,
one wants to estimate $q(x)$ stably as
$\delta \to 0$. Define
  $$ q_\delta (x):=(2\pi )^{-3}\int \limits_{|\lambda |\leq \lambda_0}
     \tilde q_\delta (\lambda )\exp (i\lambda \cdot x)\, d\lambda
     \eqno (7.3) $$
Then
  $$ \eqalign {q_\delta (x)={1\over (2\pi)^3}&\int \limits_{|\lambda |
     \leq \lambda_0}[\tilde q_\delta (\lambda )-\tilde q(\lambda )]
     \exp (i\lambda \cdot x)\, d\lambda -{1\over (2\pi )^3}\int
     \limits_{R^3\setminus B_{\lambda_0}}\tilde q(\lambda )\exp
     (i\lambda \cdot x)\, d\lambda + \cr     +{1\over (2\pi )^3}&\int
     \limits_{R^3}\tilde q(\lambda )\exp (i\lambda \cdot x)\, d\lambda
     :=I_1+I_2+q(x)}                  \eqno (7.4) $$
One has
  $$ |I_1|\leq \delta \lambda_0^3{1\over 6\pi^2}      \eqno (7.5) $$
 
  $$ |I_2|\leq {4\pi c_1\over 8\pi^3}\int \limits_{\lambda_0}^\infty
     {\lambda^2\, d\lambda \over (1+\lambda^2)^{\tilde d}}
     <{c_1\over 2\pi^2}
     \int \limits_{\lambda_0}^\infty \lambda^{-2\tilde d+2}\, d\lambda =
     {c_1\over 2\pi^2}{\lambda_0^{-2\tilde d+3}\over 2\tilde d-3}
      \eqno (7.6) $$
From (7.4)--(7.6) one gets
  $$ |q_\delta (x)-q(x)|\leq {\delta \lambda_0^3\over 6\pi^2}+
     {c_1\over 2\pi^2(2\tilde d-3)}{1\over \lambda_0^{2\tilde d-3}}
     \eqno (7.7) $$
For a fixed small $\delta $ one can minimize the right-hand side of
(7.7) in $\lambda_0$ and find $\lambda_0(\delta )$ at which the minimum
is attained:
  $$ \lambda_0 (\delta )=\left ({c_1\over \delta }\right )^{1\over 2\tilde d}
     \eqno (7.8) $$
This equation gives a practical estimate of the region on which
$\tilde q_\delta (\lambda )$ should be known for a stable recovery of
$q(x)$. With $\lambda_0=\lambda_0(\delta )$ given in (7.8) one obtains
from (7.7) the following estimate
  $$ |q_\delta (x)-q(x)|\leq c_0\delta^{1-{3\over 2\tilde d}}
  \eqno (7.9) $$
where
  $$ c_0:=\left [{1\over 6\pi^2}+{1\over 2\pi^2(2\tilde d-3)}\right ]
     c_1^{3\over 2\tilde d}         \eqno (7.10) $$
Let us formulate the result:}
\par \noindent {{\bf Lemma 21}: {\sl Let (7.1) and (7.2) hold, the
numbers $\delta $, $c_1$ and $\tilde d $
be known, $\lambda_0(\delta )$ be
defined in (7.8), and $q_\delta (x)$ be given in (7.3). Then the error
estimate of the recovery of $q(x)$ by the formula (7.3) is given by
(7.9) and (7.10).}
\par {2. Let us discuss the steps in a numerical implementation
of the methods for solving the ISP$_\delta$ developed in this paper.}
\par \noindent {{\bf Step 1}: {\sl Given $\adaa $ one calculates
$\hat A_\delta (\theta ',\alpha )$ by formula (2.25), then $u_\delta $
and $\rho_\delta (\nu )$ by formulas (2.26), (2.27) with $\theta '$
and $\theta $ satisfying (2.19).}}
\par \noindent {{\bf Step 2}: {\sl One solves the variational problem
(2.32) taking $\theta $ of order given in (2.37). This can
be done by minimizing the functional $\Vert \rho_\delta (\nu )\Vert $.
For example, one can look for $\nu $ of the form
  $$ \nu =\sum_{\ell =0}^n \nu_\ell Y_\ell (\alpha )   \eqno (7.11) $$
and find the coefficients $\nu_\ell $ from the linear system which
one gets from the condition
  $$ \Vert \rho_\delta (\nu )\Vert =\min         \eqno (7.12) $$
If $\nu_\delta (\alpha ,\theta )$ is an approximate solution to (7.12),
one checks if the inequality (2.32) holds with some constant $c$
independent of $\delta $. This is done by solving several problems
with $\delta ,{\delta \over 2},{\delta \over 4},{\delta \over 8}$ in
place of $\delta $. If the inequality (2.32) does not hold, one should
increase the number $n$ in (7.11) and try to decrease $|\theta |$.}}
\par {One may do numerical experiments with some $q(x)$, for example
$q(x)=1$ in $B_a$, $q(x)=0$ outside $B_a$, and get an idea about the
values of $\theta (\delta )$ and $\nu_\delta (\alpha )$ needed for
recovery. If formulas (2.38)--(2.39$'$) are used then only discrete
noisy data $A_\delta (\alpha_p ',\alpha_j)$ are used for recovery.}
\par {3. Let us formulate in conclusion an alternative to
(2.32) optimization method for solving the ISP with noisy data. The
{\it alternative method} consists of the following steps:}
\par \noindent {1) First, solve the problem
  $$ \Vert \rho_\delta (\nu )\Vert =\inf , \quad \nu \in L^2(S^2)
     \eqno (7.13) $$
Denote by $d(\delta ,\theta )$ the infimum in (7.13) and find
$\nu_{\theta ,\delta }(\alpha )$ such that
  $$ \Vert \rho_\delta (\nu_{\theta ,\delta })\Vert \leq
     d(\delta ,\theta )+ |\theta |^{-1}  \eqno (7.14) $$ }
\par \noindent {2) Secondly, solve the problem
  $$ H(\delta ,\theta ):=\inf :=\omega (\delta ),\quad
     \theta \in M             \eqno (7.15) $$
where
  $$ H(\delta ,\theta ):=\Vert \rho_\delta (\nu_{\theta, \delta })\Vert +
     a(\nu_{\theta ,\delta })\exp (\kappa b)\mu(\delta ),\quad
     \kappa =\hbox {Im }|\theta | ,\eqno (7.16) $$
and find $\theta_1(\delta )$ such that
  $$ H\bigl (\delta ,\theta_1(\delta )\bigr )\leq
     \omega (\delta )+\delta        \eqno (7.17) $$
Define $\hat q_{1\delta }$ by the formula
  $$ \hat q_{1\delta }=-4\pi \int \limits_{S^2}\hat A(\theta _1'
     (\delta ),\alpha )\nu_{1\delta }(\alpha )\,
     d\alpha , \quad \nu_{1\delta}:=\nu_{\theta_1(\delta ),\delta }
                     \eqno (7.18) $$
where $\theta '_1(\delta )-\theta_1(\delta )=\lambda $. Then the
following error estimate holds
  $$ \sup_{|\lambda |\leq \lambda_0}|\hat q_{1\delta }-\tilde q
     (\lambda )|\leq c[|\theta_1(\delta )|^{-1}+\delta +
     |\theta (\delta )|^{-1}]            \eqno (7.19) $$
Here $\theta (\delta )$ satisfies estimate (2.37) and
$\theta_1(\delta )$ is calculated numerically (see (7.16) and (7.17)).}
\par {Let us prove (7.19). First, note that
  $$ \Vert \rho (\nu_{1\delta})\Vert \leq c[\omega (\delta )+\delta ],
     \quad c={\rm const}\geq 1                \eqno (7.20) $$
Indeed, by (2.31),
  $$ \Vert \rho (\nu_{1\delta})\Vert \leq \Vert \rho_\delta
     (\nu_{1\delta})\Vert +ca(\nu_{1\delta }
 )\exp (\kappa_1 b)\mu(\delta )\leq
     c[\omega (\delta )+\delta ],\quad c\geq 1,\quad
     \kappa_1=|\hbox {Im }\theta_1|      \eqno (7.21) $$
Secondly, as in (6.1)--(6.3), one gets, using Lemma 6, and (3.4),
  $$ |\hat q_{1\delta }-\tilde q(\lambda )|\leq c[\Vert \rho
  (\nu_{1\delta})\Vert_a +a(\nu_{1\delta })\exp (\kappa_1b)\mu(\delta )]
\leq c[\omega (\delta )+\delta +|\theta_1(\delta )|^{-1}]\eqno (7.22) $$
Thirdly, let us prove that
  $$ \omega (\delta )\leq c|\theta (\delta )|^{-1}  \eqno (7.23) $$
where $\theta (\delta )$ is defined in (2.32) and satisfies
(2.37). One has, using (7.14),
  $$ \omega (\delta )\leq \inf_{\theta \in M}[d(\delta ,\theta )+
     |\theta |^{-1}+a(\nu_{\theta ,\delta })\exp (\kappa b)
     \mu(\delta )]\leq c|\theta (\delta )|^{-1}  \eqno (7.24) $$
where $\theta (\delta )$ satisfies (2.37).}
\par {From (7.20), (7.22) and (7.23) estimate (7.19) follows with
$|\theta (\delta )|$ satisfying (2.37). Let us summarize the result
assuming that $|\theta_1(\delta)| \to \infty$ as $\delta \to 0$}:
\par \noindent {{\bf Lemma 22}: {\sl The function (7.18) is a stable
approximation of $\tilde q(\lambda )$ and the error estimate is given
in (7.19) with $|\theta (\delta )|$ satisfying inequality (2.37) and
$|\theta_1(\delta )|$ defined by (7.17).}
\par {The result of Lemma 22 is an improvement over the result in [18].}
\par {4. The methods for solving the ${\rm ISP}_\delta $
developed in this paper are applicable in many fields. For example,
the applications to geophysical inverse problems are discussed in
[1,6,20,26], to Maxwell's equations in [1], [21], to hyperbolic
equations in [1], [22] [30],  to inverse conductivity problem
in [1],[23--25], to inverse spectral problem in [1],[29] and property C
, the basis of our theory, was introduced in [27],[6],[28] and [1]}.
\par {5. Finally, let us point out that {\it for a study of}
${\rm ISP}_\delta $ {\it it is necessary to assume that $q(x)$ vanishes
outside some ball.} Indeed, if $q(x)\in L^1(R^3)$ then the function
  $$ q_a(x)=\cases {q(x),&if $|x|>a$;\cr   0,&if $|x|<a$ \cr } $$
contributes to the scattering amplitude
  $$ \aaa =-(4\pi)^{-1}\int \limits_{R^3}\exp (-i\alpha '\cdot x)
     u(x,\alpha )q(x)\, dx  $$
the quantity which does not exceed
  $$ (4\pi)^{-1}c\int \limits_{|x|\geq a}|q(x)|\,dx \to 0 \hbox { as }
     a\to \infty , $$
where $c$ is the constant in (3.61). This contribution becomes
indistinguishable from the noise when
 $$\int \limits_{|x|\geq a}|q(x)|\,dx<4\pi c^{-1}\delta \eqno (7.25) $$
Thus, one cannot recover $q(x)$ in the region $|x|>a$, where $a$ is
determined by (7.25).}
\par {From (7.25) one can estimate the order of the radius $a$ of the
ball in which one can recover the potential given the noisy data with
the noise level $\delta $. For example, if one knows a priori that
$|q(x)|\leq c_1|x|^{-d'}$, $d'>3$, $|x|>1$, then (7.25) implies
$a^{3-d'}<((d'-3)/c_1c)\delta $, so that
  $ a>(d'-3)(c_1c)^{-1}\delta^{1/(d'-3)}$. }
\par \noindent {{\bf Acknowledgements}: The author thanks ONR, NSF
and USIEF for support, Prof. J. Sj\"ostrand whose ideas were crucial
in the proof of Lemma 6, and Dr. P. Stefanov for correspondence.
This paper was written while the author was a
Fulbright Research Professor at the Technion.}
\par \centerline {\bf REFERENCES}
\item{[0]} L. H\"ormander, {\it The Analysis of of Linear Partial
Differential Operators}, Vol. 1--4, Springer-Verlag, New York,
1983-1985.
\item{[1]} A. G. Ramm, {\it Multidemensional Inverse Scattering
Problems}, Longman, New York, 1992.(Expanded Russian edition
Mir Publishers, Moscow, 1994, pp.1-496).
\item{[2]} A. G. Ramm, {\it Random Fields Estimation Theory},
Longman, New York, 1990.(Expanded Russian edition 
Mir Publishers, Moscow, 1996).
\item{[3]} A. G. Ramm, {\it Scattering by Obstacles}, D. Reidel,
Dordrecht, 1986.
\item{[4]} A. G. Ramm, Recovery of the potential from fixed-energy
scattering data, {\sl Inverse problems \bf 4} (1988), 877--886;
{\bf 5} (1989), 255.
\item{[5]} A. G. Ramm, Stability of the numerical method for solving
3D inverse scattering problem with fixed-energy data,
{\sl Inverse Problems \bf 6} (1990), L7--L12; {\sl J. fuer die reine
und angewandte Math.},{\bf 414}, (1991),1-21.
\item{[6]} A. G. Ramm, Multidimensional inverse problems and
completeness of the products of solution to PDE,
{\sl J. Math. Anal. Appl. \bf 134} (1988), 251--253; {\bf 136} (1988),
568--574; {\bf 139} (1989), 302.
\item{[7]} A. G. Ramm, Stability of the inversion of 3D fixed-frequency
data, {\sl J. Math. Anal. Appl. bf 169} N2,(1992), 329-349.
\item{[8]} A. G. Ramm, Stability of the solution to 3D inverse
scattering problem, {\sl J. Math. Anal. Appl. bf 170} N1, (1992),1-15.
\item{[9]} A. G. Ramm, Completeness of the products of solutions to PDE
and inverse problems, {\sl Inverse Problems \bf 6} (1990), 635--664.
\item{[10]} A. G. Ramm, Property C and inverse problems, {\sl ICM-90
Satellite Conference Proceedings,} Inverse Problems in Engineering
Sciences, Proc. of a conference held in Osaka 1990, pp. 139--144.
\item{[11]} A. G. Ramm, Exact inversion of fixed-energy data, in the
book {\it Mathematical and Numerical Aspects of Wave Propagation
Phenomena}, SIAM, Philadelphia 1991, pp. 481--486.
\item{[12]} A. G. Ramm, Numerical method for solving 3D inverse problems
with complete and incomplete data, in the book {\it Wave Phenomena},
Springer-Verlag, New York 1989, pp. 34--43 (ed. L. Lam and H. Morris).
\item{[13]} A. G. Ramm, An approximation problem,
{\sl Appl. Math. Lett. \bf 4}, N5 (1991), 75--77.
\item{[14]} A. G. Ramm, Stability of the solution to the
inverse scattering problem with
exact data, {\sl Appl. Math. Lett. \bf 5}, N1 (1992), 91--94.
\item{[15]} A. G. Ramm, Numerical recovery of the 3D potential from
fixed-energy incomplete scattering data, {\sl Appl. Math. Lett. \bf 2},
N1 (1989), 101--104.
\item{[16]} A. G. Ramm, Numerical solution of 3D inverse scattering
problems with noisy discrete fixed-energy data, {\sl Appl.Math.Lett.,}
{\bf 5},N6 (1992),15-18.
\item{[17]} A. G. Ramm, Approximation by the scattering solutions and
applications to inverse scattering {\sl Math. and Comp. Modelling,
\bf 18}, N1 (1993), 47-56.
\item{[18]} A. G. Ramm, Inverse scattering with fixed-energy data
{\sl Appl.Math.Lett.,\bf 5},N4 (1992),63-67.
\item{[19]} A. G. Ramm, Spectral properties of the Schroedinger
operator in some infinite domains, {\sl Matem. Sborn. \bf 66}
(1965), 321--343.
\item{[20]} A. G. Ramm, Numerical method for solving 3D inverse
problems of geophysics, {\sl J. Math. Anal. Appl., \bf 136}
(1988), 352--356.
\item{[21]} A. G. Ramm, An inverse problem for Maxwell's equations,
{\sl Phys. Lett. \bf 138A} (1989), 459--462.
\item{[22]} A. G. Ramm and Rakesh, Property C and an inverse problem for
a hyperbolic equation, {\sl J. Math. Appl. \bf 156} (1991), 209--219.
\item{[23]} A. G. Ramm, A simple proof of uniqueness theorem in
impedance tomography, {\sl Appl. Math. Lett. \bf 1}
N3 (1988), 381--384.
\item{[24]} A. G. Ramm, Multidimensional inverse problems and
completeness of the products of homogenous PDE,
{\sl Zeitschr. f. Angew. Math. u. Mech. \bf 69}, N3 (1989), T13--T21.
\item{[25]} A. G. Ramm, Finding conductivity from boundary measurements,
{\sl Comp. and Math. with Appl. \bf 21}, N8 (1991), 85--91.
\item{[26]} A. G. Ramm, Uniqueness theorems for geophysical problems
with incomplete surface data, {\sl Appl. Math. Lett. \bf 3},
N4 (1990), 41--44.
\item{[27]} A.G.Ramm, On completeness of the products of harmonic
functions, {\sl Proc.Amer.Math.Soc.}, {\bf 99}, (1986), 253-256.
\item {[28]} A.G.Ramm, Necessary and sufficient condition for a PDO
to have property C, {\sl J. Math. Anal. Appl. \bf 156,
\sl (1991),505-509.}
\item {[29]} A.G.Ramm, Property C and uniqueness theorems for
multidimensional inverse spectral problem, {\sl Appl. Math. Lett.}
{\bf 3}, (1990),57-60.
\item {[30]} A.G.Ramm, J.Sj\"ostrand, An inverse problem for the wave
equation,{\sl Math.Zeit.},{\bf 206}, (1991), 119-130.
\item{[31]} P. Stefanov, Stability of the inverse problem in potential
scattering at fixed energy, {\sl Ann. Inst. Fourier, Grenoble \bf 40},
N4 (1990), 867--884.

$$
$$

email:    ramm@math.ksu.edu
\bye